\documentclass[prl,twocolumn,notitlepage,superscriptaddress,longbibliography]{revtex4-1}
\usepackage{amsmath}
\usepackage{amssymb}

\usepackage[unicode=true,colorlinks=true,citecolor=blue,urlcolor=blue]{hyperref}

\usepackage{bm}
\usepackage{color}
\usepackage{epsfig}

\usepackage[normalem]{ulem}

\newcommand{\blue}[1]{{\color{black}{#1}}}

\renewcommand {\phi}{{\varphi}}
\newcommand {\rmi}{{\rm i}}
\newcommand {\rmd}{{\rm d}}

\newcommand {\e}{{\rm e}}
\newcommand {\eps}{\varepsilon}

\begin{document}
\title{
Frequency combs with parity-protected cross-correlations and entanglement from dynamically modulated qubit arrays
}

\author{Denis Ilin}
\affiliation{Department of Physics and Technology, ITMO University, St. Petersburg, 197101, Russia}

\author{Alexander V. Poshakinskiy}
\affiliation{Ioffe Institute, St. Petersburg 194021, Russia}

\author{Alexander N. Poddubny}
\email{poddubny@coherent.ioffe.ru}
\affiliation{Ioffe Institute, St. Petersburg 194021, Russia}

\author{Ivan Iorsh}
\email{i.iorsh@metalab.ifmo.ru}
\affiliation{Department of Physics and Technology, ITMO University, St. Petersburg, 197101, Russia}

\begin{abstract}
We develop a general theoretical framework to dynamically engineer quantum correlations and entanglement in the frequency-comb emission from an array of superconducting qubits in a waveguide, rigorously accounting for the temporal modulation of the qubit resonance frequencies. We demonstrate, that when the resonance frequencies of the two qubits are  periodically modulated with a $\pi$ phase shift, 
it is possible to realize simultaneous bunching and antibunching in cross-correlations as well Bell states  of the scattered photons from different sidebands.  Our approach, based on the dynamical conversion between the  quantum excitations  with different parity symmetry, is quite universal. It can be used  to control \textcolor{black}{multi}-particle  correlations in generic  dynamically modulated dissipative  quantum systems.
\end{abstract}
\date{\today}

\maketitle 
\twocolumngrid

{\it Introduction.}
The ability to multiplex several signals at different frequencies and transmit them via one channel is of paramount importance for information processing.
A single photon can also be in a quantum superposition of several frequency channels and act as a flying qudit -- a multi-level analogue of the qubit -- that can be used for quantum computing~\cite{Kues2019}. 
In order to generate and process multi-qudit entanglement, one must realize (i) single qudit operations and (ii) two-qudit gates. It has been already proven theoretically and demonstrated experimentally that any single-qudit unitary operation can be performed by using a combination of phase shapers and a linear modulator~\cite{Lougovski2018}.
Realization of two-qudit gates is more complicated. The schemes with ancillas and post-selection based on the KLM protocol were proposed~\cite{Lougovski2017}. However, the use of such gates is limited since they operate only with a certain  (quite small) probability of success.
An alternative approach is to make photons interact with a quantum object with  nonlinear optical properties. As such, a two-level system (qubit) that cannot scatter two photons at once can operate as a simplest NS gate for resonant photons~\cite{Ralph2015}.  Quantum emitters with two metastable ground states
enable deterministic generation of single-rail encoded photonic cluster states~\cite{Lindner2009,Pichler2017}. 
%
In this Letter, we propose a  tunable setup with \textcolor{black}{several} waveguide-coupled qubits that realizes dynamical control of cross-correlations for multiplexed emission, enabling generation of multi-photon entangled states.  
Such states are indispensable in various areas of the emerging quantum technologies including quantum communications~\cite{marcikic2003long,gisin2002quantum} and quantum networks~\cite{monroe2014large,nguyen2019quantum,guimond2020unidirectional},  however it is rather hard to generate them using probabilistic linear-optics approaches with low success rates, and they can be vulnerable to decoherence. The scheme we put forward enables stable deterministic generation of entangled frequency-coded flying qudits.

%


\begin{figure}[b]
\centering
\includegraphics[width=0.47\textwidth]{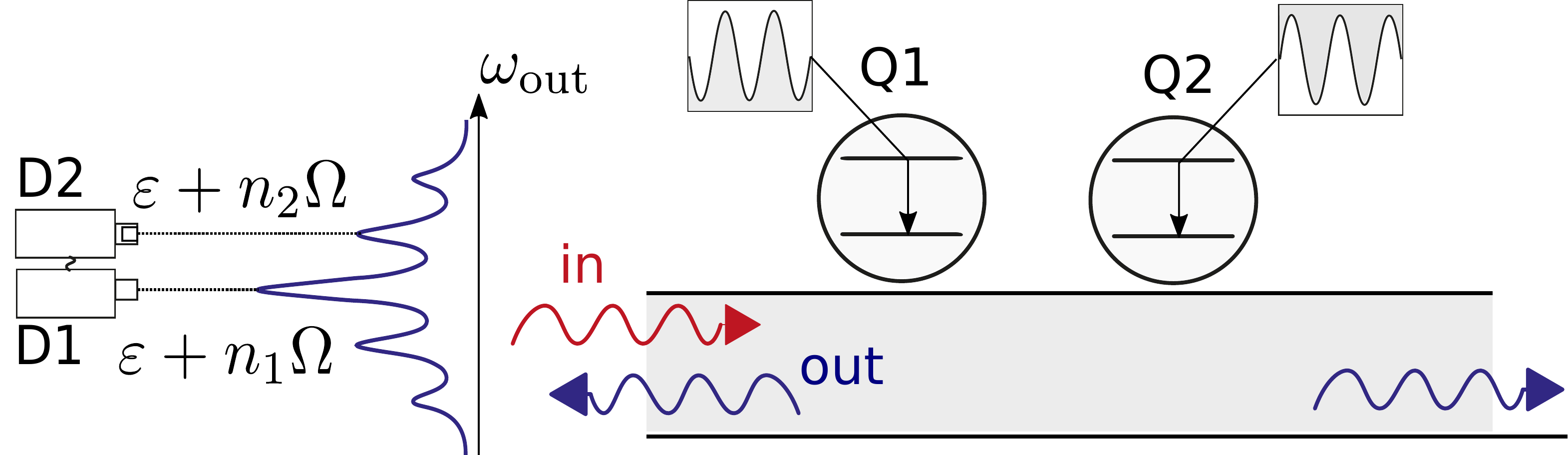}
\caption{Schematics of the structure under consideration. Two qubits Q1 and Q2 coupled to a waveguide are excited by coherent electromagnetic fields. Qubit resonance frequencies are modulated in time according to Eq.~\eqref{eq:mod}.
}\label{fig:1}
\end{figure}
%

We consider an array of qubits with the resonant frequencies harmonically modulated in time. Waveguide-coupled qubit arrays  are
now readily realized~\cite{sheremet2021waveguide}  and have  a high potential for manipulation of quantum signals~\cite{shen2007strongly,prasad2020correlating, carusotto2020photonic,kannan2020generating,Chakram2022}.
Temporal modulation can be achieved via the control optical pump beam for cold atom systems~\cite{weitenberg2021tailoring} or by means of modulated gate voltage for the case of semiconductor quantum dots or solid state defects~\cite{chen2018orbital,miao2019electrically,lukin2020spectrally,schadler2019electrical}. For  the  modulated superconducting qubits platform  the state of the art technology supports independent coherent modulation of the each individual qubit~\cite{Redchenko2022}.
We show that the qubit resonance modulation can drive the conversion between the even~(bright) and odd~(dark) states in the qubit arrays, enabling the symmetry-protected bichromatic bunching and antibunching between the photons from different sidebands. 

More generally, in sideband-resolved  regime, when the modulation frequency is much larger than the qubit resonance broadening, the frequency conversion processes, similar to Stokes and anti-Stokes Raman scattering, give rise to the frequency comb in the scattered light spectrum with multiple sidebands separated by the modulation frequency.  The correlations and entanglement  of the frequency-filtered photons in the sidebands of the emission spectrum can be quite complex. In particular, it was shown that bunched bundles of several photons can be realized by filtering certain sidebands~\cite{Laussy2020,Schmidt2021}. The advantage of our proposal is that the photon-photon correlations can be dynamically tuned, that is essential for  most of the practical applications ~\cite{jin2014ultrafast, pagliano2014dynamically}.



\begin{figure}[t]
\centering
\includegraphics[width=0.48\textwidth]{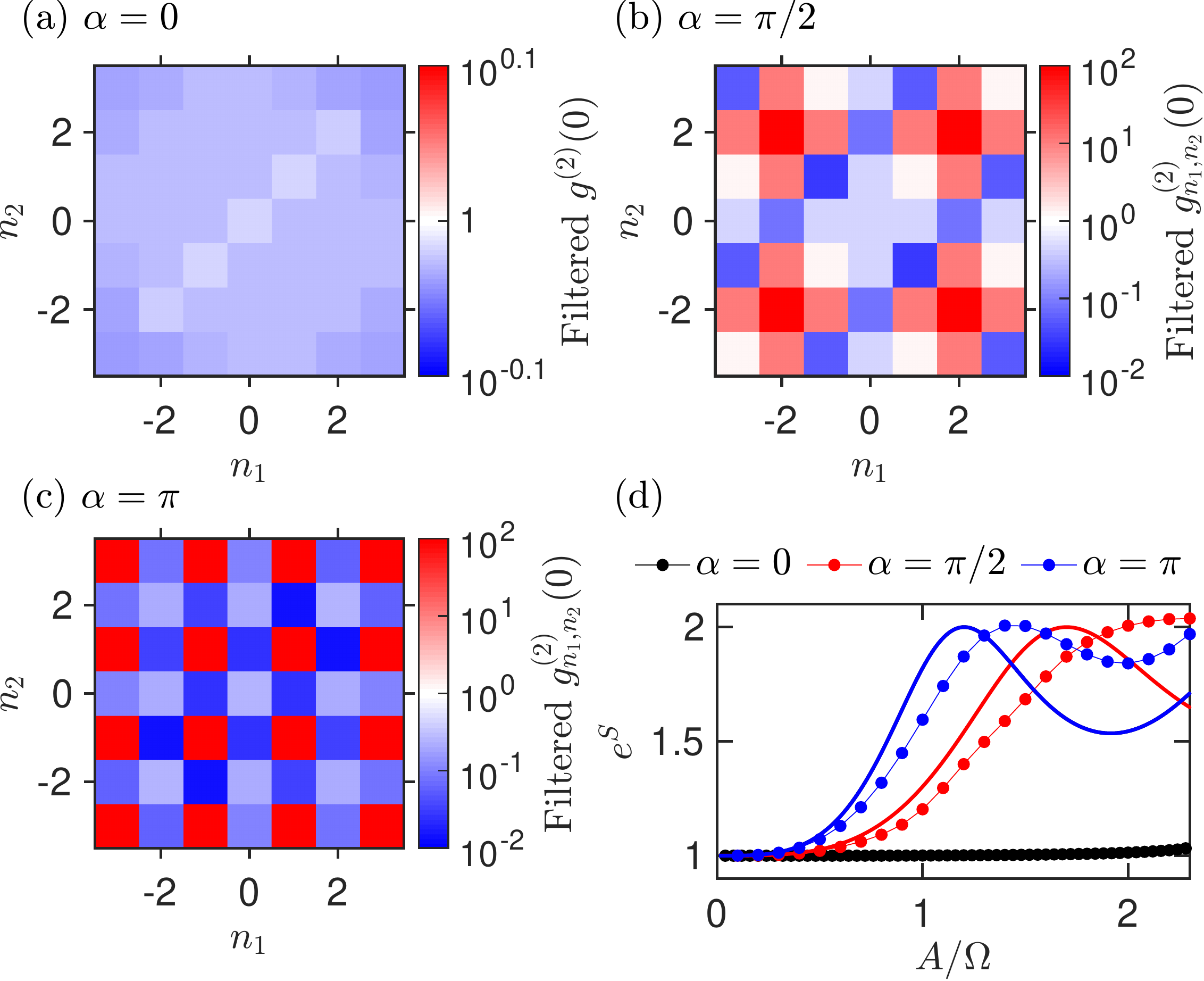}
\caption{
Bichromatic photon-photon correlations.
(a,b,c) Photon-photon correlations depending on the sideband numbers $n_1$ and $n_2$ calculated for $A=1.5\Omega$.
Other parameters are $\omega_0d/c=0$, $\Omega=200\gamma_{\rm 1D}$, $\gamma=0.05\gamma_{\rm 1D}$, $\gamma_D=5\gamma_{\rm 1D}$.  (d) Calculated exponential of the entanglement entropy $\e^S$ depending on the modulation strength $A$
for resonant excitation, $\varepsilon = \omega_0$, and the relative modulation phase $\alpha=0,\pi/2,\pi$. Thick lines show the analytical result derived in Supplementary Materials~\cite{Supp_Info1}.
}\label{fig:2}
\end{figure}

{\it Model. }The structure under consideration  consists of $N$ superconducting qubits, coupled to the waveguide, and located at the distance $d$. \textcolor{black}{We focus on the simplest case of $N=2$  qubits, shown in Fig.~\ref{fig:1} with a generalization for $N>2$ discussed in Supplementary Materials, sections (S5)-(S7)}. The qubit resonance frequencies $\omega_1$ and $\omega_2$ are modulated  as 
\begin{equation}\label{eq:mod}
    \omega_{1}(t)=\omega_0+A\cos\Omega t,\quad 
    \omega_{2}(t)=\omega_0+A\cos(\Omega t+\alpha),
\end{equation}
where $\omega_0$ is the equilibrium qubit resonance frequency, $A$ is the modulation amplitude, $\Omega$ is the modulation frequency, and $\alpha$ is the relative phase  of the modulation. The qubits are modelled as two-level systems, characterized by the spontaneous decay rate into the waveguide $\gamma_{\rm 1D}$.
The structure is excited from one side by a weak monochromatic coherent wave at frequency $\varepsilon$. We start with the sideband-resolved regime, when the modulation frequency $\Omega$ is much larger than the qubit decay rate $\gamma_{\rm 1D}$, and consider resonant excitation with frequency $\varepsilon \approx \omega_0$.  In this case, the scattered photons can have well-defined set of frequencies that form a frequency comb,
\begin{equation}\label{eq:Stokes}
    \eps+n\Omega,\quad n=0,\pm 1,\pm 2\ldots\:,
\end{equation}
where $n$ is the sideband number. Our goal is to analyze the second-order cross-correlations between the scattered photons in the sidebands $n_1$ and $n_2$
\begin{align}\label{eq:g2}
g^{(2)}_{n_1,n_2} = \frac{I^{(2)}_{n_1,n_2}}{I^{(1)}_{n_1} I^{(1)}_{n_2}} \,,
\end{align}
where $I^{(1)}_{n_{1(2)}}$ is the intensity of scattering of a single photon into sideband $n_{1(2)}$, $I^{(2)}_{n_1,n_2}$ is the intensity of scattering of a photon pair into sidebands $n_1$ and $n_2$. 



{\it Parity-protected cross-correlations.} From now on we consider the case when the two qubits are located at the same point, i.e. $\omega_0d/c=0$ (or $2\pi$), so the system is invariant under the parity operation $\mathcal P$ that interchanges  the  qubits. 
The effect of nonzero interqubit distance is analyzed in Supplementary  Materials. When such  system is not perturbed by the modulation, the light couples only to symmetric (even with respect to $\mathcal P$)  mode of the two qubits $(\sigma_1^\dag+\sigma_2^\dag)|0\rangle$, where $\sigma_{1,2}^\dag$ are the qubit raising operators. The  parity symmetry also enforces strict constraints on the photon emission of the modulated system. If the qubit modulations are in-phase, $\alpha = 0$, the photon can be scattered to any sideband. However,  for $\alpha = \pi$, the qubit energy modulation is odd with respect to $\mathcal P$. 
\blue{The  photon amplitude emitted into the sideband with even(odd) number is an even(odd) function of $A$, see Sec. 2G of the Supplementary Materials for the rigorous proof. Since it  should be invariant under $\mathcal P$, only the even-order sidebands are present in the emission spectrum.
}
Similarly, in the case of two-photon emission,  all harmonics $I^{(2)}_{n_1,n_2}$ are present if $\alpha = 0$,
but  for  $\alpha = \pi$  the $\mathcal P$ symmetry dictates that the two-photon scattering process is allowed only if $n_1+n_2$ is even. 
These symmetry arguments  indicate that the second-order cross-correlation function~\eqref{eq:g2}
should be very sensitive to the sideband numbers $n_1$ and $n_2$ when $\alpha=\pi$. In particular, for odd $n_1,n_2$ we expect  {\it parity-protected photon bunching}. 

We have modelled the two-photon frequency-filtered photon detection scheme illustrated in Fig.~\ref{fig:1}
using the master equation formalism~\cite{Blais2013},  see 
Supplementary Sec.~S3 for details. Namely, the reflected photons are absorbed by the detectors D1 and D2 and the coincidence counts are calculated~\cite{Valle2012}. The detectors are modelled as two-level systems with the frequencies $\omega_{D1}=\varepsilon+n_1\Omega$, $\omega_{D2}=\varepsilon+n_2\Omega$,
and additional nonradiative decay with the rate $\gamma_D$ that ensures  that the detectors are always well below the saturation. 
Figures~\ref{fig:2}(b--d) present the calculated  equal-time correlation function depending on the harmonic numbers $n_1$ and $n_2$ for three  relative modulation phases $\alpha=0,\pi/2,\pi$. In agreement with the symmetry analysis above, all the  harmonics are present in the emission spectrum for symmetric modulation, see Fig.~\ref{fig:2}(b). When $\alpha=\pi/2$, Fig.~\ref{fig:2}(c), the two-photon correlation pattern becomes much richer and shows alternating photon bunching and antibunching depending on the values of $n_1$ and $n_2$. This pattern is in qualitative agreement with our simplified theoretical model presented in Sec.~S4 of the Supplementary Materials. Finally, for anti-symmetric modulation, presented  in Fig.~\ref{fig:2}(d), the calculation reveals both  parity-protected photon bunching, when $n_1$ and $n_2$ are odd, and parity-protected antibunching, when $n_1$ and $n_2$ have different parity.

\begin{figure}[t]
\centering
\includegraphics[width=0.48\textwidth]{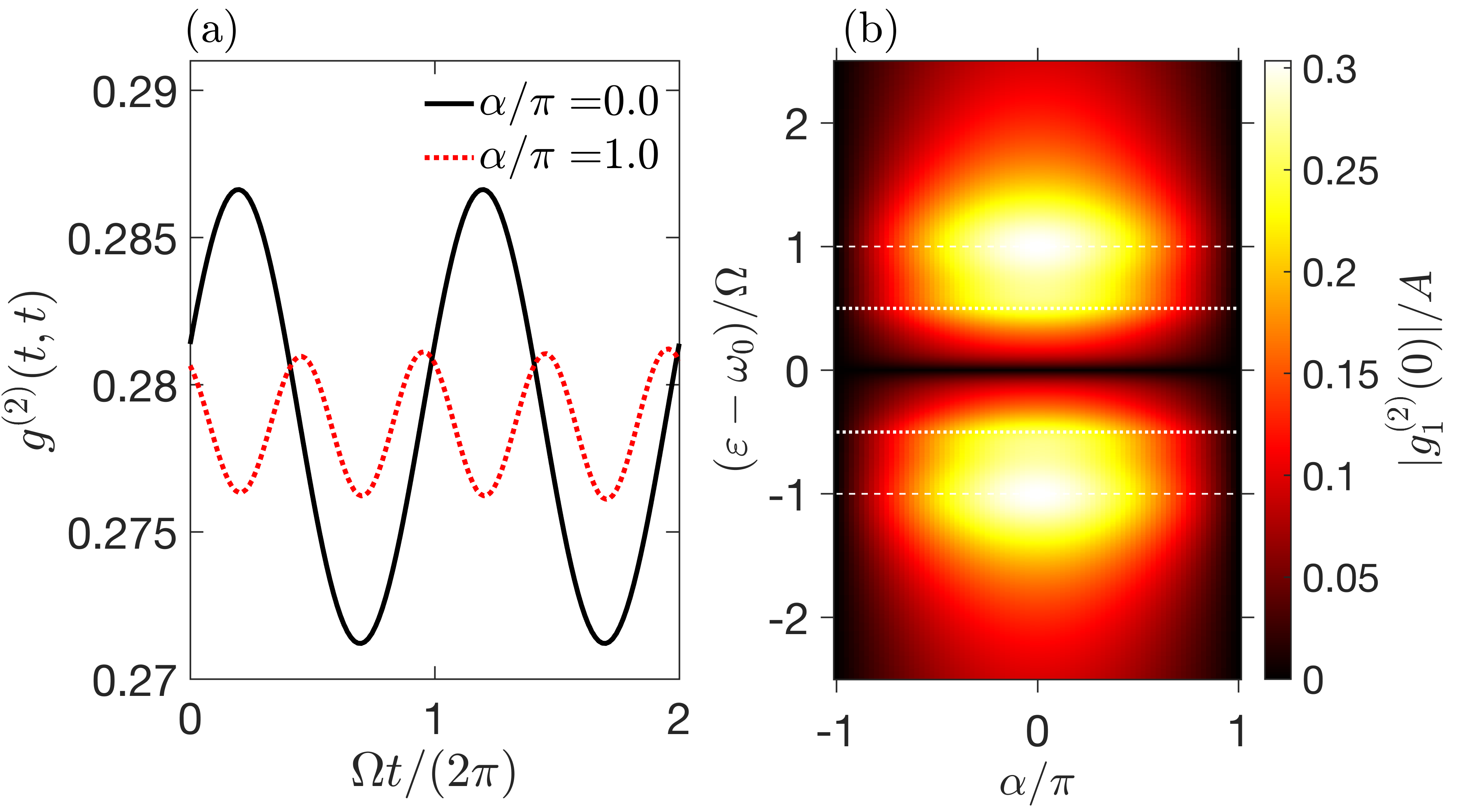}
\caption{
(a) Time-dependent photon-photon correlations $g^{(2)}(t,\tau=0)$ calculated for 
$\varepsilon-\omega_0=\Omega$. Black/solid and red/dotted curves correspond to in-phase ($\alpha=0$) and out-of-phase ($\alpha=\pi$) modulation of the first and second qubit resonance frequencies. 
(b) Color map of the correlation function first temporal harmonic $|g^{(2)}_1(0)|$ as a function of relative modulation phase $\alpha$ and frequency detuning of the incident light $\varepsilon-\omega$. The calculation parameters are $\Omega=5\gamma_{1D}$, $A=0.025\gamma_{\rm 1D}$. 
}\label{fig:3}
\end{figure}

{\it Entanglement of flying qudits.}
The emitted photons, residing in a superposition of the several frequency sidebands, can be regarded as many-level qudits. To quantify the two-qudit entanglement, we calculate the entanglement entropy \cite{Eisert2010,Poshakinskiy2021} $S =- \sum_\lambda |\lambda|^2 \ln |\lambda|^2$, where $\lambda$ is the singular value of the (normalized) two-photon wave function $\psi_{n_1,n_2}$. The latter is obtained numerically by calculating the correlation of the  detectors D1 and D2 polarizations.  The dependence of $\e^S$ on the modulation amplitude is shown in Fig.~\ref{fig:2}(d) for different relative modulation phases. For in-phase modulation, $\alpha=0$, the entropy vanishes. That follows from the rigorous analytical expression for the scattering matrix of the homogeneously modulated system that differs from that of the system without modulation only by (time-dependent) phase factors, see Supplementary Sec.~S2E. For nonzero $\alpha$, the entropy increases with $A$, reaches the maximal value of $\ln 2$, and then oscillates below it. Thick lines in Fig.~\ref{fig:2}(d) show the analytical result neglecting the radiative coupling of the qubits (see Supplementary Materials Sec.~S5). It predicts that the modulation amplitude required to achieve  $S = \ln 2$ is given by
$
    A^* = j_0\Omega/[2 \sin(\alpha/2)]
$
where $j_0$ is the zero of the Bessel function $J_0$.  Therefore, the anti-phase modulation, $\alpha = \pi$, is favorable for  maximal entanglement. 

When $S = \ln 2$, photons are in  a Bell state $\psi_{n_1,n_2} = (u_{n_1}u_{n_2}+v_{n_1}v_{n_2})/\sqrt{2}$, where $u_n$ and $v_n$ are some orthogonal single-qudit states. Using single-qudit linear operations, that can be implemented by phase shapers and optical modulators~\cite{Lougovski2018}, the Bell state can be converted to any other basis required for applications.
Another approach is to consider non-harmonic modulation of the qubits, that enables to generate any two-photon state described by a rank-2 matrix $\psi_{n_1,n_2}$, as described in the Sec.~S6 of the Supplementary Materials.

Higher rank states can be generated with  larger number of qubits $N$. \blue{As shown in Supplementary Sec. S7, the entanglement entropy of the state of $M \geq 2$ photons emitted by $N \geq M$ modulated qubits oscillates as a function of the modulation amplitude with several incommensurate periods and can reach the limiting value $\ln N$ at certain points. In particular, three qubits modulated with the amplitude $A=j_0\Omega/\sqrt{3}$ and the relative phases $2\pi/3$ emit the three-photon state $\psi_{n_1,n_2,n_3} = (u_{n_1}v_{n_2}w_{n_3} + ...)/\sqrt6$, where the ellipsis denotes the permutations of the indices, $u_n$, $v_n$, and $w_n$ are three orthogonal states. Such state possesses the maximally possible entanglement entropy of $\ln 3$ and  is an analogue of the cluster state: indeed, if one the photons is measured (in the $u$, $v$, $w$ basis), the two other photons remain in the entangled Bell state. Similar states of four and more photons can be also generated.} While the cluster states could be very important for many quantum computing applications, it is yet unclear how wide the class of many-photon states that can be generated by the proposed scheme is, e.g., if more complex matrix product states (MPS) are feasible.
 

{\it Time-dependent correlations.} \textcolor{black}{ Signatures of the sibeband cross-correlations can be observed even without frequency filtering in the time dependence of the total second order correlation function,} 
\begin{equation}\label{eq:g2t}
   g^{(2)}(t+\tau, t)=\frac{\langle a^\dag(t+\tau) a^\dag(t) a(t) a(t+\tau)\rangle }{[\langle a^\dag a \rangle_0]^2}\:,
\end{equation}
where $a$ is the annihilation operator corresponding to the reflected photons, $\langle \ldots \rangle$ and $\langle \ldots \rangle_0$ denote averaging over the state of the system with and without modulation, respectively.  
Due to the temporal modulation of the qubit resonance frequencies, the correlation function $g^{(2)}$ is no longer a function of delay time $\tau$ only, but also depends on the absolute time $t$~\cite{poshakinskiy2020} with the period $2\pi/\Omega$. This allows us to present the correlation function as the Fourier series
$g^{(2)}(t+\tau,t)=\sum_{n=-\infty}^\infty \e^{-\rmi n\Omega t}g_n^{(2)}(\tau)\:.$
We will focus on the  harmonics $g_n^{(2)}(\tau)$ at zero delay $\tau=0$. 

The numerator of the total correlation function Eq.~\eqref{eq:g2t} is determined by the squared sum 
\begin{align}\label{eq:aaaa}
\langle a^\dag(t+\tau) a^\dag(t) a(t) a(t+\tau)\rangle = \Big|\sum_n S_n(\tau)\,\e^{-\rmi\Omega t}\Big|^2 \,,
\end{align}
where $S_n$ is the amplitude of the two-photon scattering process, characterized by photon pair energy change $2\varepsilon \to 2\varepsilon + n\Omega$.
A general approach for few-photon scattering in Floquet systems was developed in~\cite{Trivedi2020}. Here we use a similar perturbative diagrammatic approach to calculate $S_n$ (see Supplementary Materials Sec.~S2).
Using Eqs.~\eqref{eq:g2t}-\eqref{eq:aaaa}, we obtain the expression 
for the $n$-th harmonic of the total two-photon correlation function,
   $ g_n^{(2)}(\tau) \propto \sum_{k=-\infty}^\infty S_{n+k}(\tau)S_k^*(\tau) \,.$
In particular, for low modulation amplitude $A$ we have $S_n \propto A^{|n|}$, so the $n=\pm 1$ harmonic $g_{1}^{(2)} \propto A$ is governed by $S_1S_0^* + S_0 S_{-1}^*$. Here $S_{\pm 1}$ correspond to  amplitudes of the first-order anti-Stokes and Stokes two-photon scattering processes $2\varepsilon \to 2\varepsilon  \pm \Omega$. 
In the considered resolved-sideband regime, they are determined by the probability of the two-photon scattering into the sidebands with  energies $\eps$ and $\eps \pm \Omega$: $I^{(2)}_{0,\pm 1} \propto |S_{\pm 1}(0)|^2$.


Our consideration of frequency-filtered correlations above has demonstrated, that for anti-symmetric modulation $\alpha=\pi$ one has $I^{(2)}_{\pm 1,0} \propto |S_{\pm 1}(0)|^2 \to 0$. Thus, we expect that  for such modulation the $n=\pm 1$ harmonic $g_{1}^{(2)}$  will be absent in the Fourier series.
This is confirmed by the rigorous calculation of the total time-dependent zero-delay correlation function $g^{(2)}(t, t)$, shown in Fig.~\ref{fig:3}(a). Black/solid and red/dotted curves correspond to $\alpha=0$ and $\alpha=\pi$, respectively. It is clearly seen from the calculation that for $\alpha=\pi$ the period of the dependence is twice smaller than that for $\alpha=0$. This indicates the absence of the first harmonic $\propto \e^{\mp\rmi\Omega t}$ in the former case and provides a direct manifestation of the parity-protected antibunching in the time-dependent photon-photon correlations.

Figure~\ref{fig:3}(b) examines the dependence of the time-resolved correlations on the relative modulation phase $\alpha$ and  the incident frequency detuning $\varepsilon-\omega$ in more detail. The color shows the numerically calculated amplitude of the first harmonic $|g^{(2)}_1(0)|/A$ (at $A \to 0$). 
In agreement with the results in Fig.~\ref{fig:3}a and Fig.~\ref{fig:2}, the correlations are suppressed if $\alpha=\pm \pi$ for any incident light frequency $\varepsilon$. The strongest correlations are achieved for the in-phase modulation, $\alpha = 0$. The calculation also shows suppression  of the harmonic $g^{(2)}_{1}(0)$ for resonant pumping, when $\varepsilon=\omega_0$. Even though the intensities of the Stokes and anti-Stokes two-photon scattering processes are nonzero in that case, $I^{(2)}_{0,\pm 1} \propto |S_{\pm 1}(0)|^2 \neq 0$, as was illustrated in Fig.~\ref{fig:2}, the interference of the two contributions to $g^{(2)}_1(0)$ stemming from the Stokes and anti-Stokes processes turns out to be destructive, $S_1(0)S_0^*(0) + S_0(0) S_{-1}^*(0) = 0$ at $\eps = \omega_0$. Note that $g^{(2)}_1(\tau)$ in this case is still nonzero if a finite delay time $\tau \neq 0$  is considered. 

As a function of pump frequency, the $g^{(2)}_{1}$ harmonic has two pairs of Stokes or anti-Stokes resonances: stronger single-photon resonances at $\eps = \omega_0 \pm \Omega$ and  weaker two-photon resonances at $2\eps = 2\omega_0 \pm \Omega$, marked by dashed and dotted lines, respectively. At  these resonances, the first harmonic of the correlation function reads
\begin{align}\label{eq:g21}
    &g^{(2)}_{1}(0) =\pm \frac{\rmi 
    A}{4\gamma_{\rm 1D}}\cos \frac{\alpha}2, \quad \eps = \omega_0 \pm \Omega \,,\\\nonumber 
    &g^{(2)}_{1}(0) = \pm \frac{7A}{6\Omega}\cos \frac{\alpha}2 ,\quad \eps = \omega_0 \pm \frac{\Omega}2
    \,,
\end{align}
where we supposed $\Omega \gg \gamma_{1D}$, see also Sec.~S3 of the Supplementary Materials for more general analytical expressions. 



\begin{figure}[t]
\centering
\includegraphics[width=0.48\textwidth]{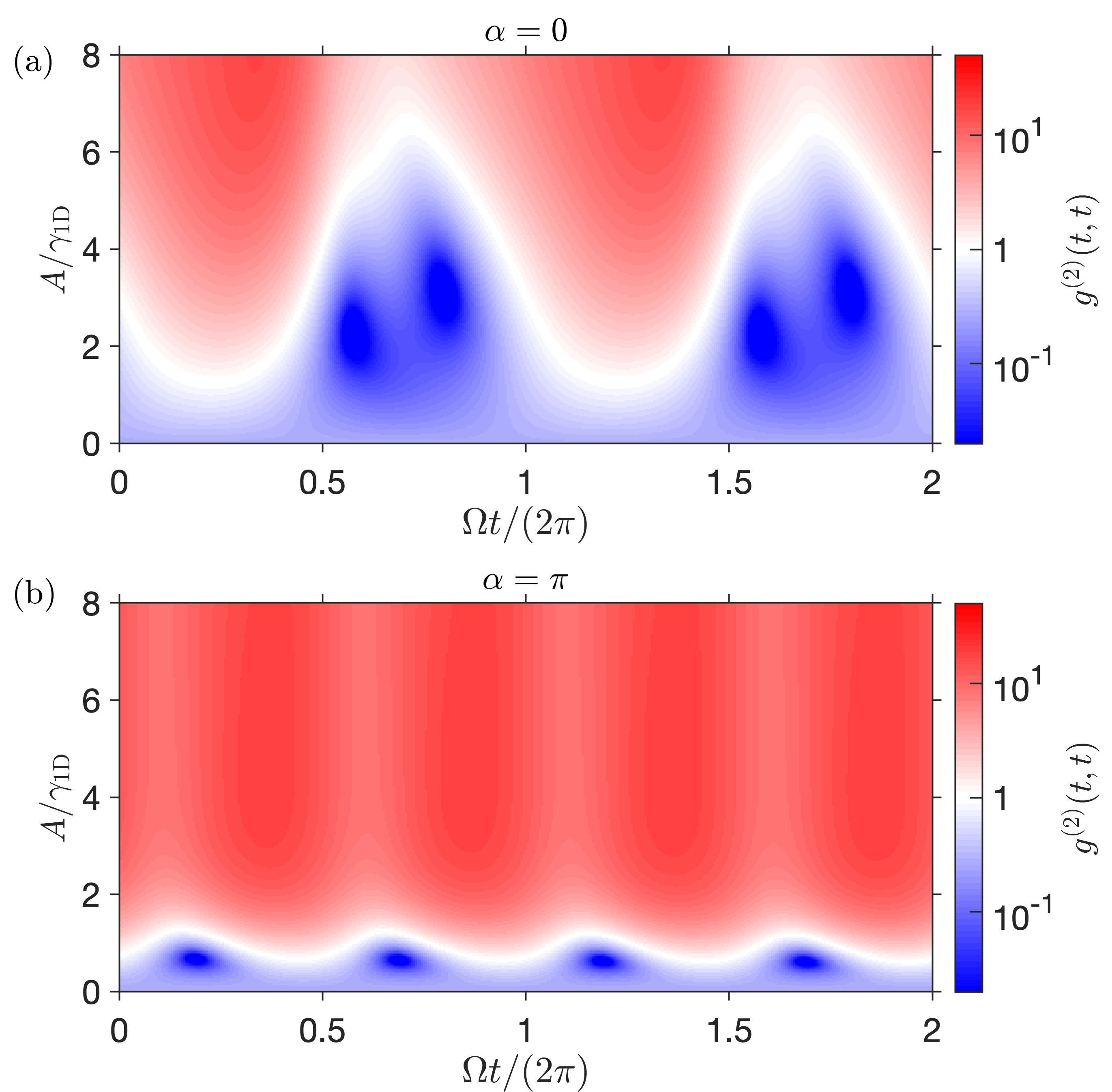}
\caption{
Total time-dependent photon-photon correlation function $g^{(2)}(t,t)$
depending on the modulation amplitude for  (a)
in-phase and (b) out-of-phase modulation of the first and second qubit resonance frequencies.
Calculation was performed for  $\eps=\omega_0+\Omega$, $\Omega=5\gamma_{1D}$.
}\label{fig:4}
\end{figure}

{\it Strong modulation.} Up to now we focused on the weak modulation case, when only the first-order Stokes and anti-Stokes scattering is considerable. For strong  modulation additional sidebands emerge, leading to  high-order harmonics $g^{(2)}_n$ in the temporal dependence of the total correlation function.
Figure~\ref{fig:4} shows the dependence of $g^{(2)}(t,t)$ on $A$ for (a) symmetric  and (b) anti-symmetric modulation. 
Similarly to the case of small $A$ (Fig.~\ref{fig:3}a),  the temporal period for anti-phase modulation is twice smaller than that for the in-phase one. This indicates the absence of all odd-order harmonics for $\alpha = \pi$,  in agreement with the parity argument forbidding  two-photon scattering processes $2\eps \to 2\eps + (2k+1)\Omega$. Then, substituting $S_{2k+1} = 0$ into $g_n^{(2)}(\tau)$, we indeed conclude that  $g^{(2)}_{2k+1} = 0$.

The correlations change significantly with the modulation strength. For low $A$, a relatively weak overall antibunching is observed for both in-phase and anti-phase modulation. With increase of $A$, the antibunching first becomes stronger, reaching maximum for $A/\gamma_{\rm 1D} \approx 3\,(0.7)$, then a bunching appears during certain time intervals, and finally the antibunching gets completely replaced by a pronounced bunching at $A/\gamma_{\rm 1D} \gtrsim 6\,(1.2)$ for the case of (anti-)symmetric modulation.
This behavior for $\alpha=0$ is well explained by Eq.~\eqref{eq:g21} that  suggests that the amplitude $|g^{(2)}_1|$ increases linearly with $A$ and should reach the value of the order of unity at the threshold $A \sim \gamma_{\rm 1D}$. Then, it can overcome the constant contribution $g^{(2)}_0$, enabling the change of the $g^{(2)}(t,t)$ sign.
Similarly, for $\alpha = \pi$ when $g^{(2)}_1 =0$, the second harmonic $g^{(2)}_2$ grows with $A$ and reaches unity at $A \sim \gamma_{\rm 1D}$.

{\it Summary.}
We have considered theoretically a waveguide QED setup where the qubit resonance frequencies are modulated periodically in time.  We predict that by tuning the relative phase of modulation  for different qubits, one can realize  \textcolor{black}{multi}-photon frequency comb in qubit emission with controllable correlations of photons at different frequencies which could be very useful in the modern optical quantum computing experiments. Our results open the way for deterministic generation and processing of entangled multi-photon states in systems with  high cooperativities, such as optical chips or chips based on superconducting qubits.

We are grateful to E.S. Redchenko for useful discussions.

%
\newpage
\onecolumngrid
\section{Supplementary information}

\tableofcontents
\section{Model}

The Hamiltonian describing an array of  oscillating quibits  in a waveguide reads
\begin{align}\label{H0}
H_0 = \sum_{n} [\omega_0+A_n(t)] \sigma^\dag_n \sigma_n + \sum_k \omega_k a_k^\dag a_k + \sum_k g \left( \sigma^\dag_n a_k \e^{\rmi k z_n} + \sigma_{n,k} a_k^\dag \e^{-\rmi k z_n} \right)
\end{align}
Here $n$ enumerates the qubits, $z_n=nd$ is the qubit coordinate, $\omega_x$ is the equilibrium qubit resonance frequency, and $A_n(t)$ is its modulation,
$\omega_k = c|k|$ is the photon dispersion in the waveguide, $g$ is the the photon-qubit interaction. We take modulation in the form
\begin{align}
 A_n(t) = u_n \e^{-\rmi\Omega t} + u_n^* \e^{\rmi\Omega t}   \,,
\end{align}
where $\Omega$ is the frequency of the modulation.  

\section{Diagrammatic approach for Stokes and anti-Stokes scattering}
In this section we outline the diagrammatic Green function approach to calculate the scattering in the first order in the modulation amplitude. The approach is conceptually similar to Ref.~\cite{dalibard1983correlation}, however, contrary to Ref.~\cite{dalibard1983correlation} it accounts for an arbitrary number of the qubits. It is instructive to replace the two-level qubits with  bosonic modes, $\sigma_n \to b_n$ and introduce  instead a Kerr nonlinearity~\cite{Poshakinskiy2016}
\begin{align}\label{eq:V}
 V =    \frac{\chi}{2} \sum_{n} b_n^\dag b_n^\dag b_n b_n\:.  \end{align}
 The two-level qubits are recovered   in the limit  $\chi \to \infty$ when the states  with two excitations residing in one qubit are excluded.

\subsection{Diagram technique}

 Diagrams describing various processes are shown in Fig.~\ref{fig:Sdia}. There, solid lines stand for qubit excitation Green's function (thin line is the bare excitation and thick line is the one dressed by waveguide photons), wavy line is the Green's function of the waveguide photon, dashed line indicates the modulation of the qubits. 
The expression for the scattering matrix element is obtained from the diagrams using the following the Feynman rules. 

\begin{enumerate}

\item There are three kind of vertices:
\begin{itemize}
\item The vertex corresponding to interaction of light with the $n$-th qubit excitation has an incoming (outgoing) photon line and outgoing (incoming) exciton line. It is associated with the factor $-\rmi g \e^{\rmi k z_n} $ ($-\rmi g \e^{-\rmi k z_n} $), where $k$ is the momentum of the involved photon.  
\item The vertex corresponding to the exciton-exciton interaction, Eq.\eqref{eq:V},  features two incoming excitons lines and two outgoing exciton lines. It is associated with the factor $-\rmi\, 2\chi$
\item The vertex corresponding to the effect of modulation that features an incoming(outgoing) dashed lines, one incoming and one outgoing exciton lines. It is associated with the factor $-\rmi\, u_n$ ($-\rmi\, u_n^*$). 

\end{itemize}

\item The vortices are connected by two types of lines:
\begin{itemize}
\item Exciton line (straight) is associated with the factor $\rmi G_{ij}$, where the bare exciton Green function reads
\begin{align}
G_{ij}^{(0)} (\omega) = \frac{\delta_{ij}}{\omega-\omega_0+\rmi 0} \,.
\end{align}
\item Photon line (wavy)  is associated with the factor $\rmi D_k$, where the bare photon Green function reads
\begin{align}
D_k (\omega) = \frac{1}{\omega-\omega_k+\rmi 0} \,.
\end{align}
\end{itemize}

\item The incident (final) photons are represented with external photon lines. They are associated with the unity factor. 

\item To obtain the scattering matrix element, 
\begin{itemize}
\item summation over all momenta and integration over all frequencies$/2\pi$ that are not fixed by conservation laws in the vertices should be performed, 
\item result should be multiplied by $2\pi\times$ $\delta$-function reflecting the energy conservation law,
\item divided by the combinatorial factor: the number of permutations of vertices  and/or lines that leave the diagram unchanged. 
\end{itemize}
\end{enumerate}

\subsection{Single-photon elastic scattering}

The Hamiltonian conserves the total number of excitations,
\begin{align}
N = \sum_k a_k^\dag a_k + \sum_j b_j^\dag b_j \,.
\end{align}
We stat with considering the sates with $N=1$. For such states, the interaction $V$ vanishes. 

\begin{figure}[t!]
\includegraphics[width=0.99\columnwidth]{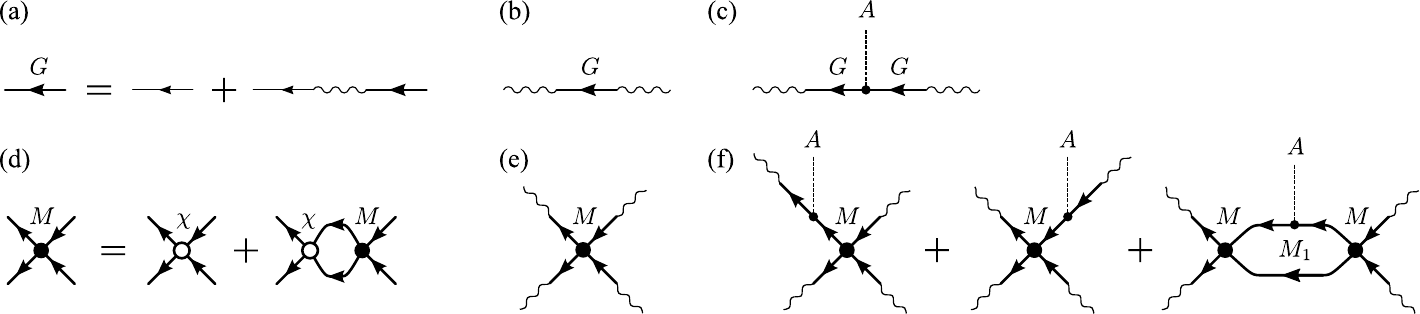}
\caption{
Diagrammatic representation of photon scattering on modulated qubits. (a) Dyson equation for the Green's function of qubit excitation. Thin solid line is the bare qubit excitation, wavy line is the photon in the waveguide, thick solid line is the qubit excitation dressed by interaction with photons. (b) Diagram describing elastic scattering of a single photon. (c) Diagram describing inelastic scattering of a single photon with emission/absorption of a modulation quantum (dashed line). (d) Dyson equation for a pair of qubit excitations. Open dot denotes bare vertex of excitation interaction, Eq.~\ref{eq:V}. Solid dot is the dressed interaction vertex, Eq.~\eqref{eq:M}.
(b) Diagram describing scattering of a photon pair without change of the total energy. (c) Diagrams describing inelastic scattering of a photon pair  with emission/absorption of a modulation quantum.
} \label{fig:Sdia}
\end{figure}

First, we account for the interaction of the qubit excitations with photons. The Dyson-like equation, describing the dressing of qubit excitations, is shown in Fig.~\ref{fig:Sdia}(a) and reads
\begin{align}\label{eq:G11}
{G}_{ij}(\omega) = {G}_{ij}^{(0)}(\omega) + \sum_k \sum_{l,m} {G}_{il}^{(0)}(\omega)\, g_k \e^{\rmi k z_l}\, D_k(\omega)\,g_k  \e^{-\rmi k z_m} \,{G}_{mj}(\omega) \,.
\end{align}
Summation over $k$ can be easily performed assuming linear dispersion $\omega_k = c |k|$ and constant $g$,
\begin{align}
\sum_k \frac{g_k^2 \e^{\rmi k (z_l-z_m)}}{\omega-\omega_k + \rmi 0} = -\rmi\, \frac{g^2}{v}\, \e^{\rmi(\omega/c)|z_l-z_m|} \,.
\end{align}
Then, Eq.~\eqref{eq:G11} assumes the form
\begin{equation}  \label{eq:defG}
(\omega-\omega_0) {G}_{ij}(\omega)+\rmi\gamma_{\rm 1D}\sum_m \e^{\rmi (\omega/c)|z_i-z_m|} {G}_{mj}(\omega)= \delta_{ij}\:,
\end{equation}
where $\gamma_{\rm 1D}=g^2/c$ is the rate of spontaneous emission into the waveguide. In other words,
 the dressed
 matrix Green's function of the qubit excitations reads
\begin{align}
\bm G(\omega) = (\omega-\bm H)^{-1}
\end{align}
where
\begin{align}\label{H1}
H_{ij} = \omega_0 \delta_{ij} - \rmi\gamma_{1D} \e^{\rmi q |z_i-z_j|}
\end{align}
is the effective non-Hermitian Hamiltonian, accounting for qubit-photon interaction with the traced out the photonic degrees of freedom~ \cite{Caneva_2015,Ke2019},  and $q=\omega/c$ is the photon wave vector.  In what follows, we work in the Markovian approximation which corresponds to using $q=\omega_0/c$s,  that is valid provided that $\gamma_{\rm 1D}\ll \omega_0$.

The amplitude of elastic photon reflection  is given by the diagram in Fig.~\ref{fig:Sdia}(b) and reads
\begin{align}
r(\omega_k) = \langle a_{k}  | S | a_k^\dag \rangle =  -\rmi \gamma_{1D} \sum_{ij} G_{ij} \e^{iq(z_i+z_j)} \,.
\end{align}

\subsection{Single-photon anti-Stokes scattering}

The amplitude of photon reflection with absorption of a single vibration quantum $r_1$ is defined from the scattering matrix $S$ as
\begin{align}
\langle a_{k'}  | S | a_k^\dag \rangle = r_1(\omega_k)\,  \frac{2\pi c}{L} \delta (\omega_{k'}+\Omega - \omega_k) \:,
\end{align}
where $L$ is the normalization length.
It is described by the diagram in Fig.~\ref{fig:Sdia}(c), where the dashed line indicates the modulation, and can be 
calculated as 
\begin{align}\label{eq:r1gen}
r_1(\omega) = \gamma_{1D} \sum_{ijk} G_{ik}(\omega+\Omega) u_k G_{kj}(\omega) \e^{iq(z_i+z_j)} = \gamma_{1D} \sum_{k} u_k s^+_k(\omega+\Omega)  s^+_k(\omega) \,,
\end{align}
where 
\begin{align}
s^+_i(\omega) = \sum_j G_{ij}(\omega) \e^{iqz_j}.
\end{align}

Equations for the scattering amplitude can be simplified for the homogeneous modulation, $u_k = u ={\rm const}$.
In the time domain, such the modulation leads just to appearance of the (time-dependent) phase factor for the single-photon scattering matrix 
\begin{equation}\label{eq:Stt}
S(t',t) = S_0(t',t)\,\e^{-\rmi\int_t^{t'} 2u\cos\Omega \tau \rmd \tau}\equiv
S_0(t'-t)\, \e^{-\rmi\frac{2u}{\Omega}[\sin\Omega t'-\sin\Omega t]}
\end{equation}
Switching back to the frequency domain, we get
\begin{equation}
\label{eq:S1b}
S(\omega+n\Omega\leftarrow \omega)=
\sum_{k=-\infty}^{\infty} J_{n+k}\left(\frac{2u}{\Omega}\right) r(\omega-k\Omega) \,
 J_{k}\left(\frac{2u}{\Omega}\right) \,
\end{equation}
where $J$ is the Bessel function. 
Equation~\eqref{eq:S1b} is valid for an arbitrary ratio  $u/\Omega$. In the limit $u\ll\Omega$, we extract the amplitude of first-order scattering process: 
\begin{equation}\label{eq:r1b}
r_{1}(\omega)=S_1(\omega+\Omega\leftarrow \omega)= 
\frac{u}{\Omega}[r(\omega)-r(\omega+\Omega)]\:.
\end{equation}

\subsection{Two-photon elastic scattering}
In order to describe the two-photon scattering, we use the approach from Ref.~\cite{Ke2019}. First, we introduce the dressed vertex $M_{ij}$ that describes interaction of two qubit excitations. From the the diagrammatic equation shown in the  Fig.~\ref{fig:Sdia}(d) we find
\begin{align}\label{eq:M}
(M^{-1})_{ij}(\epsilon) = -\rmi \int G_{ij}(\omega)G_{ij}(2\epsilon-\omega) \frac{d\omega}{2\pi}  
= - \left( \frac1{2\epsilon - H \otimes I - I \otimes H}\right)_{ii,jj} \,,
\end{align}
where $I$ is the identity matrix. 
The elastic two-photon scattering amplitude $S_0$ defined by 
\begin{align}
\langle a_{k_1'} a_{k_2'}  | S | a_{k_1}^\dag a_{k_2}^\dag\rangle = 2\pi \delta (\omega_{k_1'} +\omega_{k_2'}  -\omega_{k_1} - \omega_{k_2}) \, S_0(\omega_{k_1'} ,\omega_{k_2'} ; \omega_{k_1} , \omega_{k_2}) \,.
\end{align}
The incoherent contributions to $S_0$ are by the diagram in Fig.~\ref{fig:Sdia}(e) and yield
\begin{align}
S_0(\omega_1',\omega_2';\omega_1,\omega_2) &= 2\pi [\delta(\omega_1'-\omega_1)+\delta(\omega_1'-\omega_2)] t(\omega_1) t(\omega_2)\\
& - 2 \rmi \gamma_{1D}^2 \sum_{ij} M_{ij}(\tfrac{\omega_1+\omega_2}2) s^+_i(\omega_1') s^+_i(\omega_2') s^+_j(\omega_1) s^+_j(\omega_2) \,.
\end{align}

\subsection{Two-photon anti-Stokes scattering}

The anti-Stokes two-photon scattering amplitude $S_1$ is defined by
\begin{align}
\langle a_{k_1'} a_{k_2'}  | S | a_{k_1}^\dag a_{k_2}^\dag\rangle = 2\pi \delta (\omega_{k_1'} +\omega_{k_2'}  -\omega_{k_1} - \omega_{k_2}-\Omega) \, S_1(\omega_{k_1'} ,\omega_{k_2'} ; \omega_{k_1} , \omega_{k_2}) \,.
\end{align}
The diagrams contributing to the incoherent part of $S_1$ are shown in Fig.~\ref{fig:Sdia}(f) and yield
\begin{align}
S_1&(\omega_1',\omega_2';\omega_1,\omega_2) = 2\pi[\delta(\omega_1'-\omega_1)+ \delta(\omega_2'-\omega_1)] r(\omega_1) r_1(\omega_2) \\\nonumber
&\hspace{3cm}+  2\pi[\delta(\omega_1'-\omega_2)+ \delta(\omega_2'-\omega_2)] r(\omega_2) r_1(\omega_1) \\\nonumber
&+2\gamma_{1D}^2 \sum_{ijk} u_k  \Big\{ M_{ij}(\epsilon)  \big[ s^+_k(\omega_1') G_{ki}(\omega_1'-\Omega) s^+_i(\omega_2') + s^+_i(\omega_1') G_{ki}(\omega_2'-\Omega) s^+_k(\omega_2') \big]  s^+_j(\omega_1) s^+_j(\omega_2) \\\nonumber
&\hspace{1cm}+M_{ij}(\epsilon+\frac{\Omega}{2})s^+_i(\omega_1') s^+_i(\omega_2') \big[ s^+_k(\omega_1) G_{kj}(\omega_1+\Omega) s^+_j(\omega_2) + s^+_j(\omega_1) G_{kj}(\omega_2+\Omega) s^+_k(\omega_2) \big]  \Big\} \\\nonumber
%
&+ 4\gamma_{1D}^2 \sum_{ijkl} M_{ik}(\varepsilon+\tfrac{\Omega}2) M_{1,kl}(\epsilon) M_{lj}(\varepsilon) s^+_i(\omega_1') s^+_i(\omega_2') s^+_j(\omega_1) s^+_j(\omega_2) \,,
\end{align}
where
\begin{align}
M_{1,ij}(\epsilon) &= \rmi \int \sum_k  u_k G_{ij}(\omega)G_{ik}(2\epsilon-\omega+\Omega)G_{kj}(2\epsilon-\omega) \frac{d\omega}{2\pi}
\\ \nonumber
& = \left( \frac1{2\epsilon +\Omega - H \otimes I - I \otimes H} \,\text{diag}(u) \otimes I \,\frac1{2\epsilon - H \otimes I - I \otimes H}\right)_{ii,jj} \,.
\end{align}

In case of homogeneous modulation, $u_k = u$, the scattering  matrix $S_1$ can be expressed via $S_0$ in a similar fashion to Eq.~\eqref{eq:S1b}:
\begin{align}
    S(t_1',t_2';t_1,t_2) = S_0(t_1',t_2';t_1,t_2) \,
    \e^{-\rmi\frac{2u}{\Omega}[\sin\Omega t_1'+\sin\Omega t_2'-\sin\Omega t_1-\sin\Omega t_2']} \,.
\end{align}
Switching back to the frequency domain, we get
\begin{align}\label{eq:Sjjjj}
S(\omega_1',\omega_2'; \omega_1,\omega_2)=
\sum_{k_1',k_2',k_1,k_2}^{\infty} J_{k_1'}\left(\frac{2u}{\Omega}\right) 
J_{k_2'}\left(\frac{2u}{\Omega}\right) 
J_{k_1}\left(\frac{2u}{\Omega}\right) 
J_{k_2}\left(\frac{2u}{\Omega}\right) \nonumber\\
S_0(\omega_1'-k_1'\Omega,\omega_2'-k_2'\Omega; \omega_1-k_1\Omega,\omega_2-k_2\Omega)
\,.
\end{align}
Considering the limit of small $u$, we get
\begin{align}
&S_1(\omega_1',\omega_2'; \omega_1,\omega_2)=\\\nonumber
&\frac{u}{\Omega}[S(\omega_1'-\Omega,\omega_2'; \omega_1,\omega_2)
+S(\omega_1',\omega_2'-\Omega; \omega_1,\omega_2)
-S(\omega_1',\omega_2'; \omega_1+\Omega,\omega_2)
-S(\omega_1',\omega_2'; \omega_1,\omega_2+\Omega)] \,.
\end{align}

\subsection{Cross-correlations}
The wavefunction of the system can be expanded in series
$\psi=\psi_0+\psi_1\e^{-\rmi\Omega t}+\ldots$. Here, the term $\psi_0$ is modulation-independent  and
determined by the Fourier transform of the scattering matrix $S_0$. The term $\psi_1$ is linear in modulation and is determined by the Fourier transform of $S_1$.
We define  the time-resolved cross-correlation function of the photons in the 0-th and 1-st sideband as
\begin{equation}
g^{(2)}_{0,1}(\tau)=\frac{|\langle \psi_1|a^\dag(0) a^\dag(\tau)a(\tau)a(0) |\psi_1\rangle |^2}{8|r_\omega|^2|r_{1,\omega}|^2}\:.\label{eq:defg21}
\end{equation}
Equation~\eqref{eq:defg21} is normalized in such way that it does not depend on the modulation amplitude.
In the  case of excitation with energy $\eps$ the correlation function reads
\begin{align}\label{eq:g21gen}
g^{(2)}_{0,1}(\tau) = \frac{|\int S_1(\omega,2\eps+\Omega-\omega;\eps,\eps)\,\e^{-\rmi\omega \tau} \, \rmd\omega/(2\pi)|^2}{8 |r(\eps)|^2 |r_1(\eps)|^2} \,.
\end{align}

In the general case, Eq.~\eqref{eq:g21gen} can be evaluated numerically using Cauchy theorem for integration. For a single qubit $N=1$, the result reads
\begin{equation}
g^{(2)}_{1}(\tau) =\frac1{2}\left|\e^{-\rmi\Delta \tau}+\e^{-\rmi ( \Delta+\Omega) \tau}-
{\frac { ( \Omega+\Delta+\Omega+\rmi\gamma_{\rm 1D}) {\e^{- \gamma_{\rm 1D}\tau-\rmi \Omega \tau}}+ \left( \Omega-\Delta-\rmi\gamma_{\rm 1D} \right) {{\rm e}^{-\gamma_{\rm 1D}\tau}}}{\Omega}}
\right|^2,
\end{equation}
where $\Delta=\eps-\omega_0$.

\blue{
\subsection{The proof of the selection rules for two qubits}

In principle, diagrammatic approach can be used to calculate the amplitude of higher-order sidebands. To describe the process with the change of the photon energy (or the energy of photon pair in case of two-photon scattering) by $n \Omega$ one should sum up the diagrams that have $n_+$ incoming and $n_-$ outgoing dashed lines (modulation vertices) related by $n_+ - n_- = n$. The amplitude corresponding to such diagrams  is proportional to $A^m$ where $m = n_+ + n_-$. Note that $m$ and $n$ have the same parity. Therefore, the amplitude of a photon or photon pair emission into the sibeband with even(odd) number $n$ is an even(odd) function of the modulation amplitude $A$. 

In case of two qubits with anti-symmetric energy modulation, the transformation $A \to -A$ corresponds to the swap of the qubits and must not change the emission amplitude, since the qubits are located in the same point. Therefore, only sidebands with even numbers are allowed.
}

\section{Density matrix approach for real-time evolution of correlations}

For large modulation amplitudes, the correlation functions may comprise many high-order Stokes and anti-Stokes scattering processes. Instead of the summation over all of them, we develop here an alternative approach based of the real-time evolution of the density matrix. 

We use the master equation~\cite{Blais2013}
\begin{eqnarray}\label{eq:difsystem}
   \dot{\rho}=-i[H_1,\rho]+\sum_{j,k=1}^N\gamma_{\rm 1D}\cos[q(z_j-z_k)]\left[2\sigma_j\rho\sigma_k^{\dagger}-\{ \sigma_k^{\dagger}\sigma_j ,\rho\}\right].
\end{eqnarray}
with the Hamiltonian
\begin{eqnarray}\label{eq:H2}
    H_1&=    \sum\limits_{j=1}^N[\omega_0+A_j(t)]
    \sigma_j^\dag \sigma_j+\gamma_{\rm 1D} \sum\limits_{j,k=1}^N\sigma_j^\dag \sigma_k\sin(q |z_j-z_k|)\\\nonumber
    &- \sum\limits_{j=1}^N\frac{\rmi \Omega_R}{2} (\e^{-\rmi q z_j-\rmi \varepsilon t}\sigma_j^\dag -{\rm H.c.})
\end{eqnarray}
The first line in Eq.~\eqref{eq:H2} presents the real part of the qubit Hamiltonian~\eqref{H1} and the second line accounts for the coherent excitation at the frequency $\varepsilon$ with the strength determined by the Rabi frequency $\Omega_R$. The imaginary part of the Hamiltonian~\eqref{H1} is accounted by the Lindblad operator, last term in Eq.~\eqref{eq:difsystem}. This master equation is valid in the Markovian approximation when the flight time of light between the qubits is  small.

We are interested in the time dependence of the correlation function
\begin{equation}\label{eq:g2t}
   g^{(2)}(t+\tau, t)=\frac{\langle a^\dag(t+\tau) a^\dag(t) a(t) a(t+\tau)\rangle }{[\langle a^\dag a \rangle_0]^2}\:,
\end{equation}
where $a$ is the annihilation operator corresponding to the reflected photons that is found as
\begin{eqnarray}
    a(t)=\rmi[\sigma_{1}(t)+\e^{\rmi\phi}\sigma_{2}(t)]\:,
\end{eqnarray}
 $\langle \ldots \rangle$ and $\langle \ldots \rangle_0$ denote averaging over the state of the system with and without modulation, respectively. Equation~\eqref{eq:g2t} is normalized to the squared photon number calculated neglecting the effect modulation, 
$[\langle a^\dag a \rangle_0]^2$.  The correlation function can be readily evaluated by solving numerically the master equation and using the quantum regression theorem~\cite{Carmichael}.

However, it is instructive to find an analytical solution for the case of small modulation amplitude $A\ll \gamma_{\rm 1D}$ and weak driving strength $\Omega_R\ll \gamma_{\rm 1D}$. For $N=2$ qubits located at the same point, $\varphi=0$,
the result reads
\begin{equation}\label{eq:g20}
    g^{(2)}(t,t)=g^{(2)}_0+g^{(2)}_1e^{-\rmi\Omega t}+g^{(2)}_{-1}e^{\rmi\Omega t}
\end{equation}
where
\begin{align}
    \label{eq:g21}
            g^{(2)}_0&=\frac{1+(\Delta/2)^2}{1+\Delta^2}\:,\\\nonumber
    g^{(2)}_{-1}&={g^{(2)}_{1}}^*=\displaystyle-g^{(2)}_0\frac{2A\Delta}{\gamma_{\rm 1D}}  \cos\frac{\alpha}{2}\frac{10+4\Delta^2+7\rmi\Omega/\gamma_{\rm 1D}-\Omega^2/\gamma_{\rm 1D}^2}{[(2\Delta)^2-(-2\rmi+\Omega/\gamma_{\rm 1D})^2][\Delta^2-(-2\rmi+\Omega/\gamma_{\rm 1D})^2]}\:,
\end{align}
where $\alpha$ is the relative modulation phase and $\Delta= (\eps-\omega_0)/\gamma_{1D}$ is the pump frequency detuning. 
Eq.~\eqref{eq:g21} shows that  the $g^{(2)}_{1}$ harmonic has two pairs of Stokes or anti-Stokes resonances: stronger single-photon resonances at the driving frequency $\eps = \omega_0 \pm \Omega$ and a weaker two-photon resonances at $2\eps = 2\omega_0 \pm \Omega$. Simplifying Eq.~\eqref{eq:g21} near these resonances we obtain Eq.~(8) from the main text. 


\subsection{Effect of nonzero inter-qubit distance}

We now consider the situation when the qubits are spatially separated by some distance $d$. We analyzed how the finite phase $\phi = \omega_0 d/c$, that is gained by light while travelling between the qubits, affects our results.

\begin{figure}[h]
\centering
\includegraphics[width=0.5\textwidth]{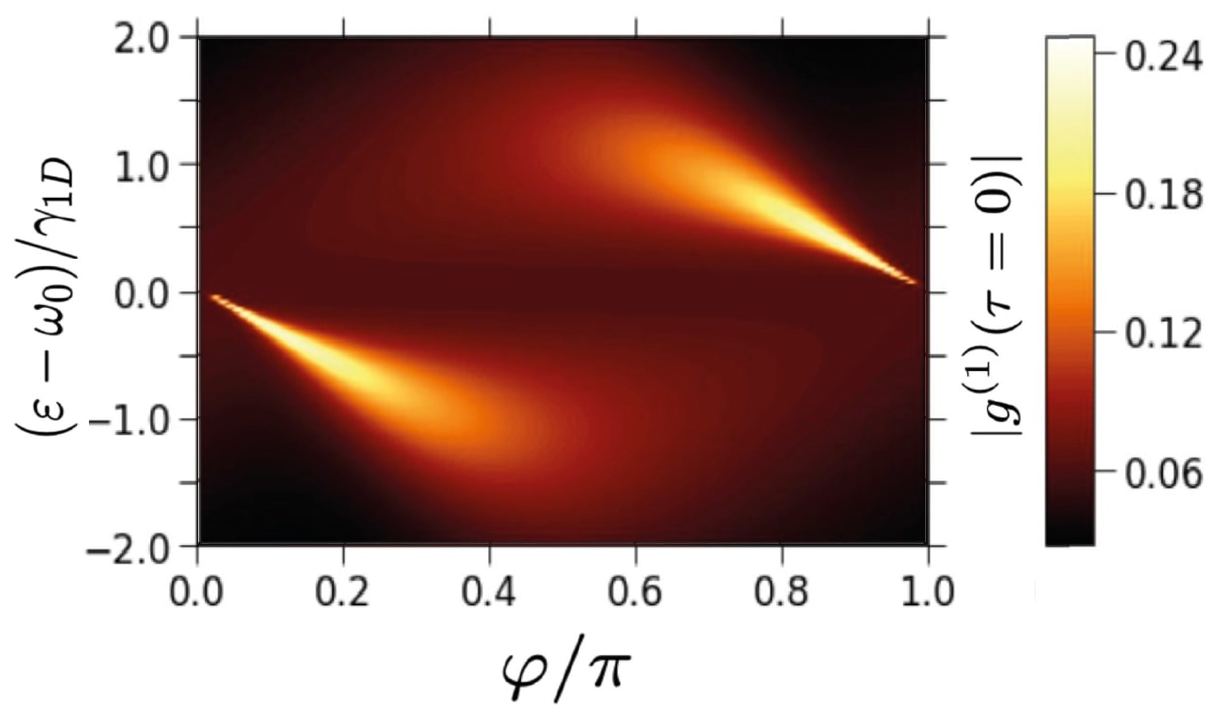}
\caption{The first-order correlation function $g^{(1)}$ of the light coherently reflected from a pair of qubits as a function of the detuning $(\varepsilon-\omega_0)/\gamma_{1D}$ and the inter-qubit distance $\phi$. Calculation is done in the  absence of the modulation.
} \label{fig:s2}
\end{figure}

 Fig.~\ref{fig:s2} illustrates the amplitude of coherent  reflection in the absence of modulation, calculated as a function of excitation frequency $\eps$ and the inter-qubit distance. The calculations reveals two resonances that shift and change their width with $\phi$. 
 In the vicinity of Bragg resonances, $\phi =0, \pi$, the wide resonance corresponds to the superradiant (bright) mode with the large decay rate $ 2\gamma_{\rm 1D}$, and the other narrow resonance corresponds to the subradiant (dark) with almost zero decay rate. In the anti-Bragg case, $\phi=\pi/2$, both modes are bright with the same radiative decay $\gamma_{1D}$, but their energies are split by $2\gamma_{\rm 1D}$. 

\begin{figure}[h]
\centering
\includegraphics[width=0.9\textwidth]{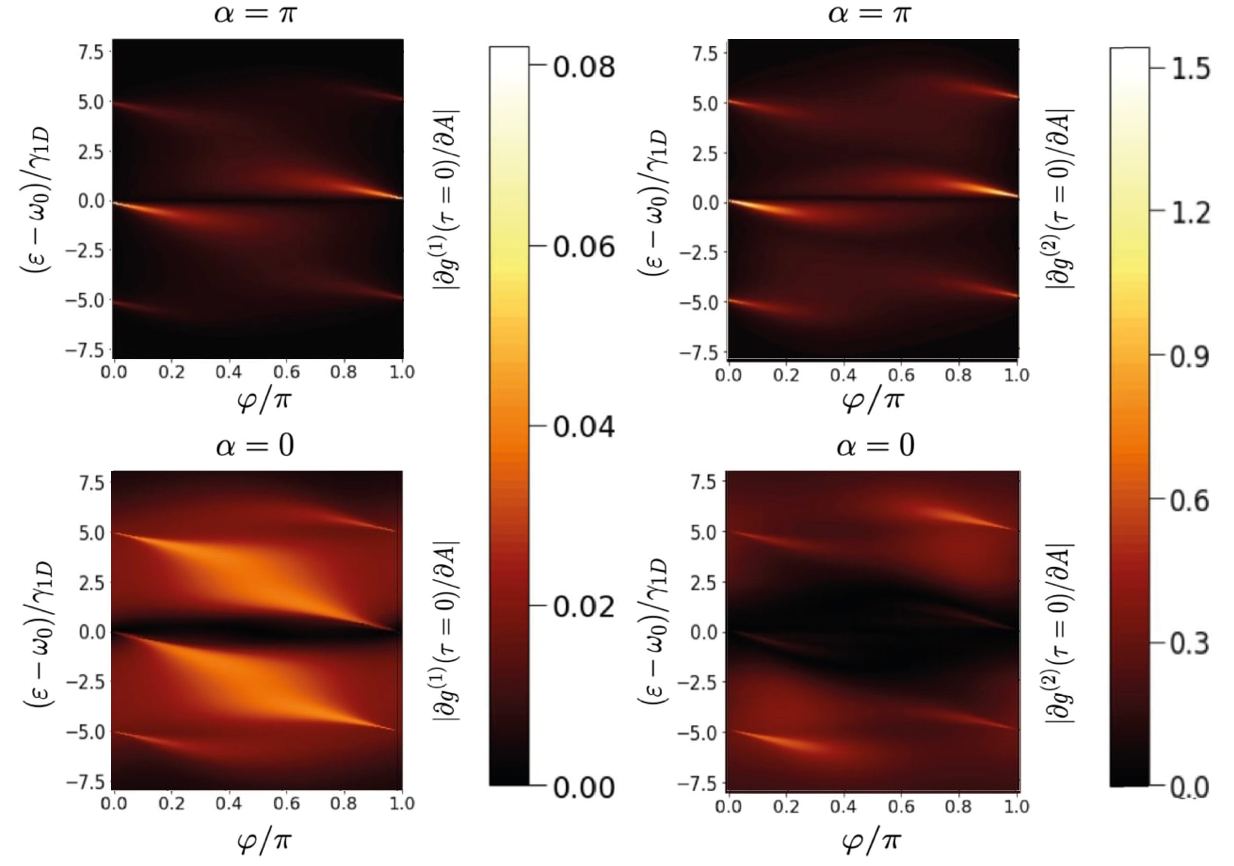}
\caption{
The first- and second-order correlation functions $|\partial g^{(1)}(\tau=0)/\partial A|$  and $|\partial g^{(2)}(\tau=0)/\partial A|$ for the light  inelastically  reflected from a pair of qubits as a function of the detuning $(\varepsilon-\omega_0)/\gamma_{1D}$ and the inter-qubit distance $\phi$. Calculation is performed for the cases of anti-symmetric ($\alpha=\pi$) and symmetric ($\alpha=0$) modulation with the frequency $\Omega/\Gamma=5$. The obtained values for $\alpha=\pi$ case were reduced by a factor of 40 in the left column and by a factor 15 in the right column.}
\label{fig:s3}
\end{figure}

Figure~\ref{fig:s3} shows how  finite $\phi$ affects the single and two-photon inelastic reflection in the presence of modulation with the relative phase $\alpha=0,\pi$. 
In the left and right panels we show by color the cross-correlation fucntions $|\partial g^{(1)}(\tau=0)/\partial A|$  and $|\partial g^{(2)}(\tau=0)/\partial A|$, which quantify the amplitude of single- and two-photons first-order inelastic scattering, respectively.  The scattering amplitudes demonstrate three pairs of resonances, corresponding to the Rayleigh, Stocks and anti-Stocks cases when $\eps$, $\eps-\Omega$, or $\eps+\Omega$ match the frequencies of the single-photon eigenmodes. Note that for anti-symmetric modulation ($\alpha=\pi$, upper panels), the wide superradiant mode does not contribute to the scattering amplitude. That is the consequence of parity symmetry, as described in the main text.  


\subsection{Frequency-filtered photon-photon correlations}
In order to calculate the 
frequency-filtered photon-photon correlations, shown in Fig.~2 of the main text, we add two additional qubits  \#3 and \#4 to the  system with the frequencies 
\begin{equation}
    \omega_{\rm D1}=\eps+n_1\Omega,\quad 
    \omega_{\rm D2}=\eps+n_2\Omega,\quad n_{1,2}=0,\pm 1,\pm 2\ldots
\end{equation}  that serve as detectors of the reflected photons  in the sidebands $n_1$ and $n_2$.
As a result, there are $N=4$ qubits in the system in total.
The master equation Eq.~\eqref{eq:difsystem} is  modified to \begin{eqnarray}\label{eq:difsystem2}
   \dot{\rho}=-\rmi[\widetilde H_1,\rho]+\sum_{j,k=1}^N\gamma_{j,k}\left[2\sigma_j\rho\sigma_k^{\dagger}-\{ \sigma_k^{\dagger}\sigma_j ,\rho\}\right].
\end{eqnarray}
where 
\begin{equation}
    \gamma_{j,k}=\gamma_{\rm 1D}\cos[q(z_j-z_k)]+\gamma_D(\delta_{j,2}+\delta_{j,3})\delta_{j,k}\:.
\end{equation}
Here we have added the fast decay term $\gamma_{\rm D}\gg \gamma_{\rm 1D}$ to the detectors in order to ensure that their population is kept low and that the reemission of the absorbed photons from the detectors  is negligible.
The modified Hamiltonian reads
\begin{multline}\label{eq:H2b}
    \widetilde H_1=    \sum\limits_{j=1}^2[\omega_0+A_n(t)]+\omega_{\rm D1}\sigma_3^\dag \sigma_3+\omega_{\rm D2}\sigma_4^\dag \sigma_4\\+
    \gamma_{\rm 1D} \sum\limits_{j,k=1}^N\sigma_j^\dag \sigma_k\sin(q |z_j-z_k|)-\frac{\rmi \Omega_R}{2} \sum\limits_{j=1}^2 (\e^{-\rmi q z_j-\rmi \varepsilon t}\sigma_j^\dag -{\rm H.c.})
\end{multline}
The  detectors are placed at the left from the first qubit. We solve numerically the master equation for four qubits and calculate the frequency-filtered photon-photon correlation function as 
\begin{equation}
    g^{(2)}_{n_1,n_2}=\frac{\langle \sigma_3^\dag \sigma_4^\dag  \sigma_4\sigma_3 \rangle}{\langle \sigma_3^\dag \sigma_3 \rangle \langle \sigma_4^\dag \sigma_4 \rangle}\:.
\end{equation}

\section{Selection rules for photon-photon correlations}\label{sec:simple}

\begin{figure}[t]
\label{max}
\centering
\includegraphics[width=0.75\textwidth]{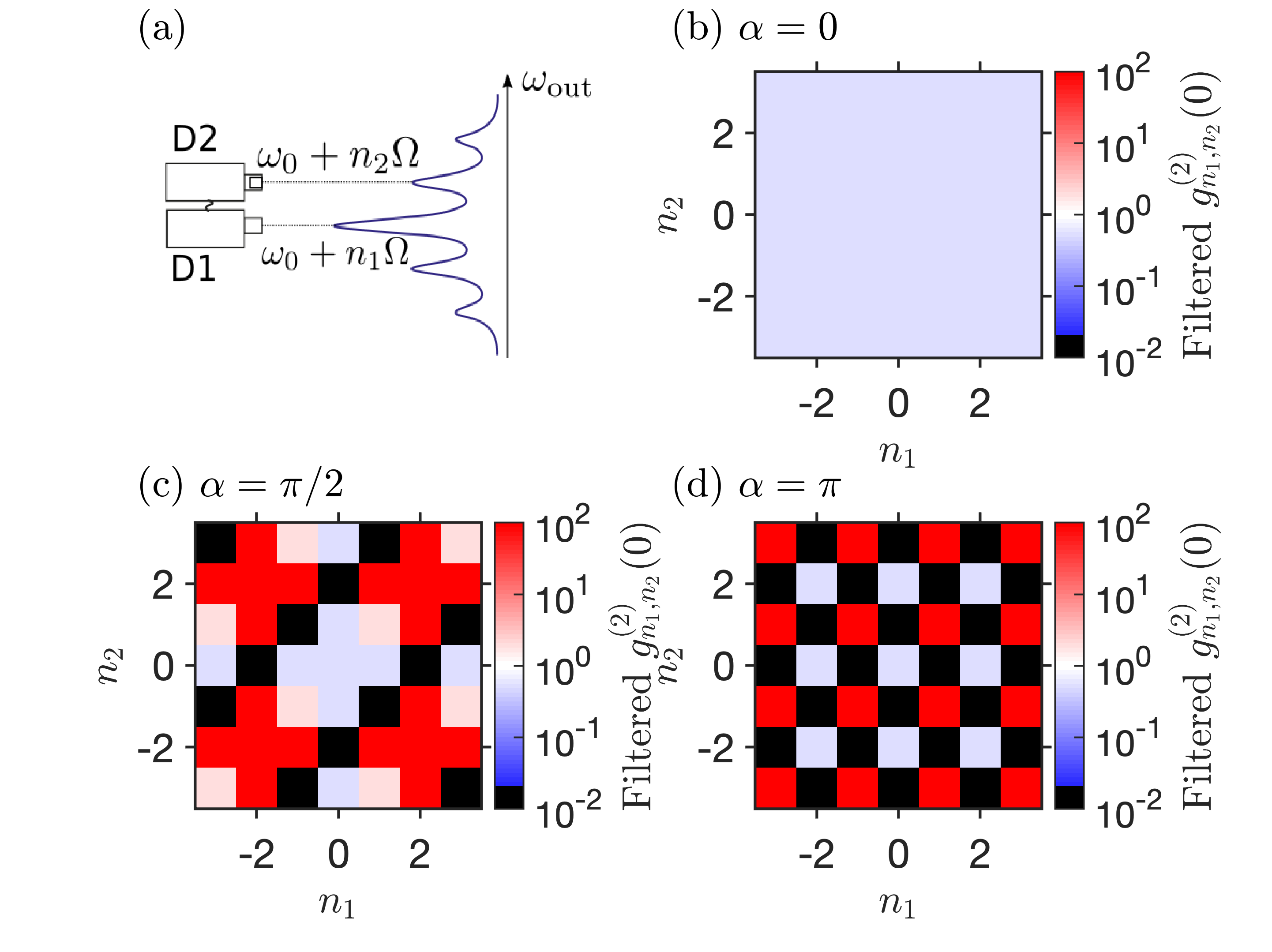}
\caption{
Cross-correlations of the spectrally filtered photons depending on the detected harmonic numbers.
Same as Fig.~2 in the main text, but calculated analytically according to Eqs.~\eqref{eq:I1}-\eqref{eq:g2}.
(a) Schematics of the measurement protocol with  detectors D1 and D2 filtering the frequencies $\omega_0+n_1\Omega$, $\omega_0+n_2\Omega$.
(b,c,d) Correlations for modulation of the qubit resonance frequencies depending on the detection parameters $n_1$ and $n_2$ calculated for the relative phase $\alpha=0,\pi/2,\pi$.
Black color corresponds to the points where Eq.~\eqref{eq:g2} is undefined. Deep red color corresponds to $g^{(2)}_{n_1,n_2}\to \infty$.
}\label{fig:4}
\end{figure}


{ Here, we present a simple approach to estimate the cross-correlations between the scattered photons.
Neglecting the radiative decay and coupling between the qubits, the evolution of the $\sigma_{j}$ operator in the Heisenberg picture reads
\begin{align}\label{eq:supSigN}
    \sigma_{j}(t) = \sigma_{j}(0) \,\e^{-\rmi \int_0^t \omega_j(t')\, dt'} = \sum_n \sigma_{j}^{(n)} \, \e^{-\rmi (\omega_0+n \Omega) t} \,.
\end{align}
Here, we introduced operators $\sigma_{j}^{(n)}$ which correspond to the $n$-th sideband. Using Eq.~(1) for $\omega_j(t')$ and evaluating the integral in Eq.~\eqref{eq:supSigN} we obtain
\begin{align}
  &\sigma_{1}^{(n)} = J_n(\tfrac{A}{\Omega})\,\sigma_{1}(0) \,,\\
  &\sigma_{2}^{(n)} = J_n(\tfrac{A}{\Omega})\,\e^
  {-\rmi n \alpha}\,\sigma_{2}(0) \,.
\end{align}
The light emission to the $n$-th sideband (in the case $\phi =0$) is determined by 
\begin{align}
\sigma^{(n)} = \sigma_{1}^{(n)}+\sigma_{2}^{(n)} \,.
\end{align}
}

We start by considering  the scattering of a single photon. When it is absorbed, the system resides in the symmetric superposition $|\psi_1\rangle = (\sigma_1^\dag + \sigma_2^\dag)|0\rangle$. The intensity of the photon emission in the $n$-th sideband is calculated as 
\begin{align}\label{eq:I1}
I^{(n)}_1 \propto |\langle 0| \sigma^{(n)} |\psi_1\rangle|^2 = 2 J_n^2(\tfrac{A}{\Omega})\, (1+\cos n\alpha) \,.
\end{align}
Importantly,  if the qubit modulations are in-phase, $\alpha = 0$, the photon can be scattered to any sideband, while in case of anti-phase modulation, $\alpha = \pi$, only even sidebands are allowed. The latter result is the consequence of the parity symmetry. Indeed, the unperturbed system is invariant under the operation $\mathcal P$ that interchanges  the two qubits. For $\alpha = \pi$, the qubit energy modulation is odd with respect to $\mathcal P$. Since the emitted light amplitude is even under $\mathcal P$, only the even powers of $A$ can contribute to it, meaning only even-order sidebands are present.

We now consider the two-photon scattering that provides insight about the second-order photon-photon correlations. After the absorption of two photons, the system lands in the only double-excited state $|\psi_2\rangle = \sigma_1^\dag \sigma_2^\dag|0\rangle$ present in the system (we recall that a qubit cannot be excited twice). Then, the probability of the emission of two photons into the sidebands $n_1$ and $n_2$ reads
\begin{align}\label{eq:I2}
I^{(n_1,n_2)}_2 \propto  \, |\langle 0| \sigma^{(n_1)} \sigma^{(n_2)} |\psi_2\rangle|^2 
= 2 J_{n_1}^2(\tfrac{A}{\Omega})J_{n_2}^2(\tfrac{A}{\Omega})\, [1+\cos (n_1-n_2)\alpha] \,.
\end{align}
Similarly to the single-photon case, all harmonics are present if $\alpha = 0$. In case   $\alpha = \pi$ the two-photon scattering process is allowed only if $n_1-n_2$ is even, which is also a consequence of the $\mathcal P$ symmetry. 

The cross-correlation function of the scattered light in sidebands $n_1$ and $n_2$ is defined as
\begin{align}\label{eq:g2}
g^{(2)}_{n_1,n_2} = \frac{I^{(n_1,n_2)}_2}{I^{(n_1)}_1 I^{(n_2)}_1} \,.
\end{align}
If $\phi=\pi$ and $n_1$, $n_2$ both odd, it follows from Eqs.~\eqref{eq:I1}-\eqref{eq:I2} that $I^{(n_1)}_1  = I^{(n_2)}_1 =0$ while $I^{(n_1,n_2)}_2$ is finite, so we get strong bunching $g^{(2)}_{n_1,n_2}  \to \infty$ protected by the parity symmetry. If $n_1$ and $n_2$ are both even, the  $g^{(2)}_{n_1,n_2}  $ is finite and determined by the detuning of photon energies from the qubit resonance. Finally if $n_1$ is even and $n_2$ is odd, Eq.~\eqref{eq:g2} is indeterminate, since both 
$I^{(n_1,n_2)}_2$ and $I^{(n_2)}_1$ turn zero.


Figure~\ref{fig:4} shows the cross-correlation functions calculated after Eqs.~\eqref{eq:I1}-\eqref{eq:g2} for different $\alpha$. They reveal the same pattern as does the rigorous calculation presented in the Fig. 2 of the main text.


\section{Entanglement entropy}

\begin{figure}[t]
\label{max}
\centering\includegraphics[width=0.4\textwidth]{ 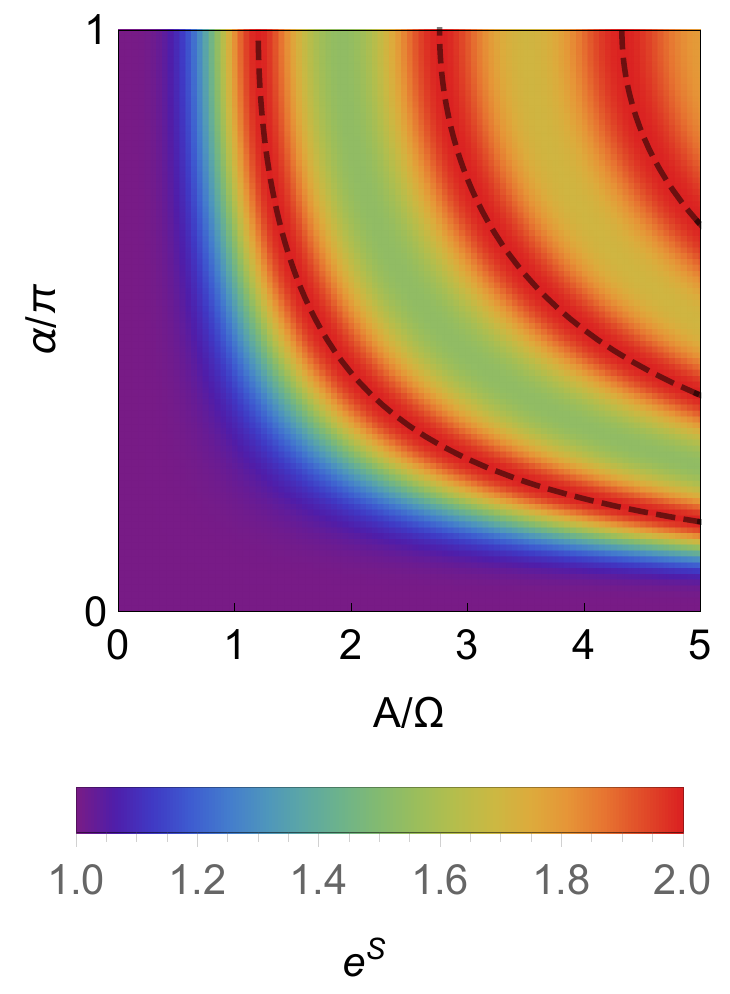}%
\includegraphics[width=0.6\textwidth]{ 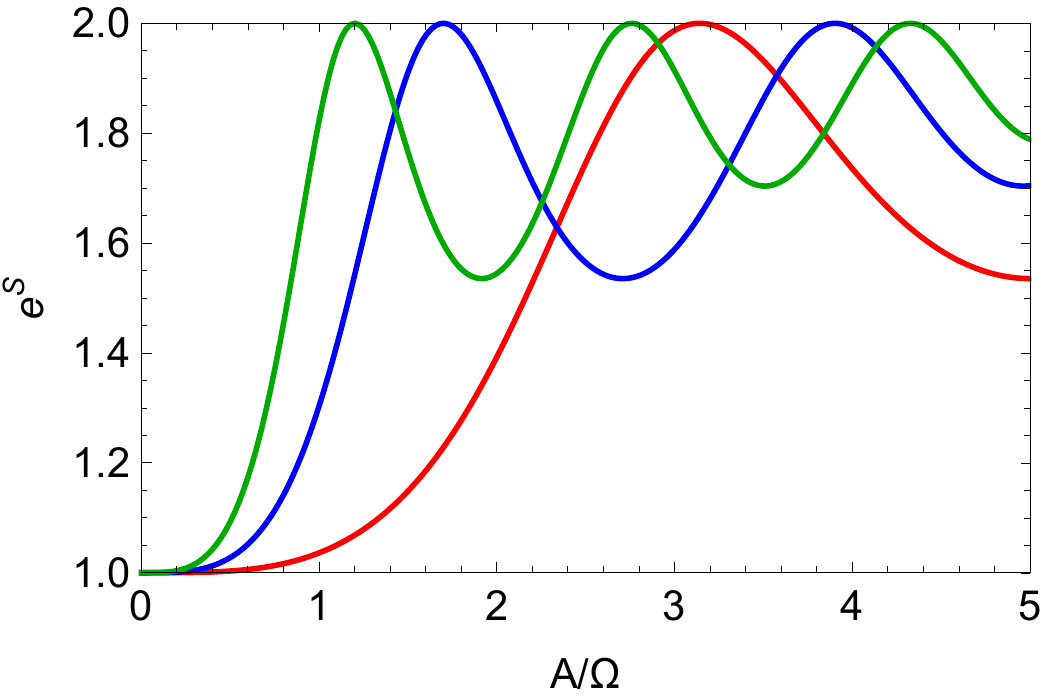}
\caption{(left) Color plot of $\e^S$, where $S$ is the entanglement entropy, for the pair of photons reflected from two modulated qubits as a function of the modulation amplitude $A$ and the relative phase $\alpha$. (right) Cross-sections  at $\alpha=\pi/4$, $\pi/2$, and $\pi$ (red, blue, and green lines). 
}\label{fig:entr2}
\end{figure}


The emitted photons can be regarded as qudits -- many-level quantum systems. Indeed, each photon can reside in one of many sidebands, or in their superposition. The effective Hilbert space dimension of such frequency qudits can be estimated as $\sim 2A/\Omega$, since the sidebands with numbers $|n| \gtrsim A/\Omega$ are weakly excited due to quenching of the Bessel function $J_n(A/\Omega)$. 

The quantum state of a pair of such flying qudits is described by the wave function $\psi_{n_1n_2}$, that depends on the sideband number of the two emitted photons $n_1$ and $n_2$. 
To quantify the entanglement of the flying qudits, we calculate the entanglement entropy $S$ \cite{Eisert2010} defined as
\begin{align}\label{eq:defentr}
    S =- \sum_\nu |\lambda_\nu|^2 \ln |\lambda_\nu|^2 \,
\end{align}
where $\lambda_\nu$ are the singular values of the matrix $\psi_{n_1n_2}$ that should be normalized according to $\sum_\nu |\lambda_\nu|^2 =1$. We note that if all the qubits are modulated in-phase, the rigorous analytical answer for the scattering
matrix is given by Eq.~\eqref{eq:Sjjjj}. For the resonant excitation and resolved-sideband regime, the scattered
photon pair wave function reads 
\begin{align}
    \psi_{n_1,n_2} \propto J_{n_1}(A/\Omega)J_{n_2}(A/\Omega) \,.
\end{align}
Since $\psi_{n_1,n_2}$ factorizes with respect to $n_1$ and $n_2$, the entanglement entropy is equal to zero. Therefore, at least two qubits modulated with different phases must be considered in order to get an entangled state. 

For two qubits, the wave function of the emitted photon pair at zero delay time can be obtained using the simplified approach of the previous section:
\begin{align}\label{eq:psinn}
   \psi_{n_1,n_2} =  a_{n_1}^{(1)} a_{n_2}^{(2)} + a_{n_1}^{(2)} a_{n_2}^{(1)}
\end{align}
where $a_n^{(i)}$ is defined from
\begin{align}\label{eq:an}
    \sum_n a_n^{(i)} \exp(-i n \Omega t) = \exp\left( -\rmi\int_0^{t'} \omega^{(i)}(t')dt' \right) \,.
\end{align}
For harmonic modulation with equal amplitudes $A$ and the phase delay $\alpha$, this yields
\begin{align}\label{eq:psinnH}
    \psi_{n_1,n_2} \propto J_{n_1}(A/\Omega)J_{n_2}(A/\Omega) \left( \e^{-\rmi n_1 \alpha} +\e^{-\rmi n_2 \alpha}\right) \,.
\end{align}
We note that both of the two terms in Eq.~\eqref{eq:psinn} are factorized with respect to $n_1$ and $n_2$. Therefore, the entanglement entropy cannot exceed $\ln 2$. Straightforward  calculation shows that the two nonzero singular values,
\begin{align}
    |\lambda_{1,2}| = \frac{1\pm x}{\sqrt{2(1+x^2)}} \,, 
\end{align}
are defined by a scalar product of the frequency combs generated by the two qubits:
\begin{align}\label{eq:scal}
    x &=(a^{(1)},a^{(2)}) = \sum_n a^{(1)*}_n a^{(2)}_n
    =\int_0^{2\pi/\Omega} \exp\left\{\rmi \int_0^t \left[\omega_1(t')-\omega_2(t')\right]\, \rmd t'\right\}  \frac{\Omega\,\rmd t}{2\pi} \nonumber\\
    &=J_0\left(\frac{2A}{\Omega} \, \sin \frac{\alpha}{2} \right) \,.
\end{align}

Figure~\ref{fig:entr2} shows the calculated entanglement entropy as a function of modulation amplitude $A$ and relative phase $\alpha$. If either $A$ or $\alpha$ is zero, the entropy vanishes. The maximal value of entropy $S=\ln 2$ is reached on the dashed lines described by
\begin{align} \label{eq:entr2}
  J_0\left(\frac{2A}{\Omega} \, \sin \frac{\alpha}{2} \right) = 0 \,,  
\end{align}
which corresponds to the case when frequency combs generated by the two qubits are orthogonal. In such case, the two singular values of $\psi_{n_1,n_2}$ are equal to  $1/\sqrt 2$. The wave function of the photon pair reduces to the Bell state
\begin{align}
    \psi_{n_1,n_2} = \frac1{\sqrt{2}}(u_{n_1}u_{n_2} + v_{n_1}v_{n_2})
\end{align}
in a certain orthogonal basis basis $(u, v)$. While this basis is rather complicated and mixes several sidebands, using pulse shapers and linear modulators, it can be changed to virtually any other basis that is more convenient for the practical use of the Bell state. It follows from Eq.\eqref{eq:entr2} that anti-phase modulation, $\alpha = \pi$, is preferable for achieving maximal entanglement at smaller modulation amplitude.

\begin{figure}[t]
\label{max}
\centering
\includegraphics[width=0.99\textwidth]{ 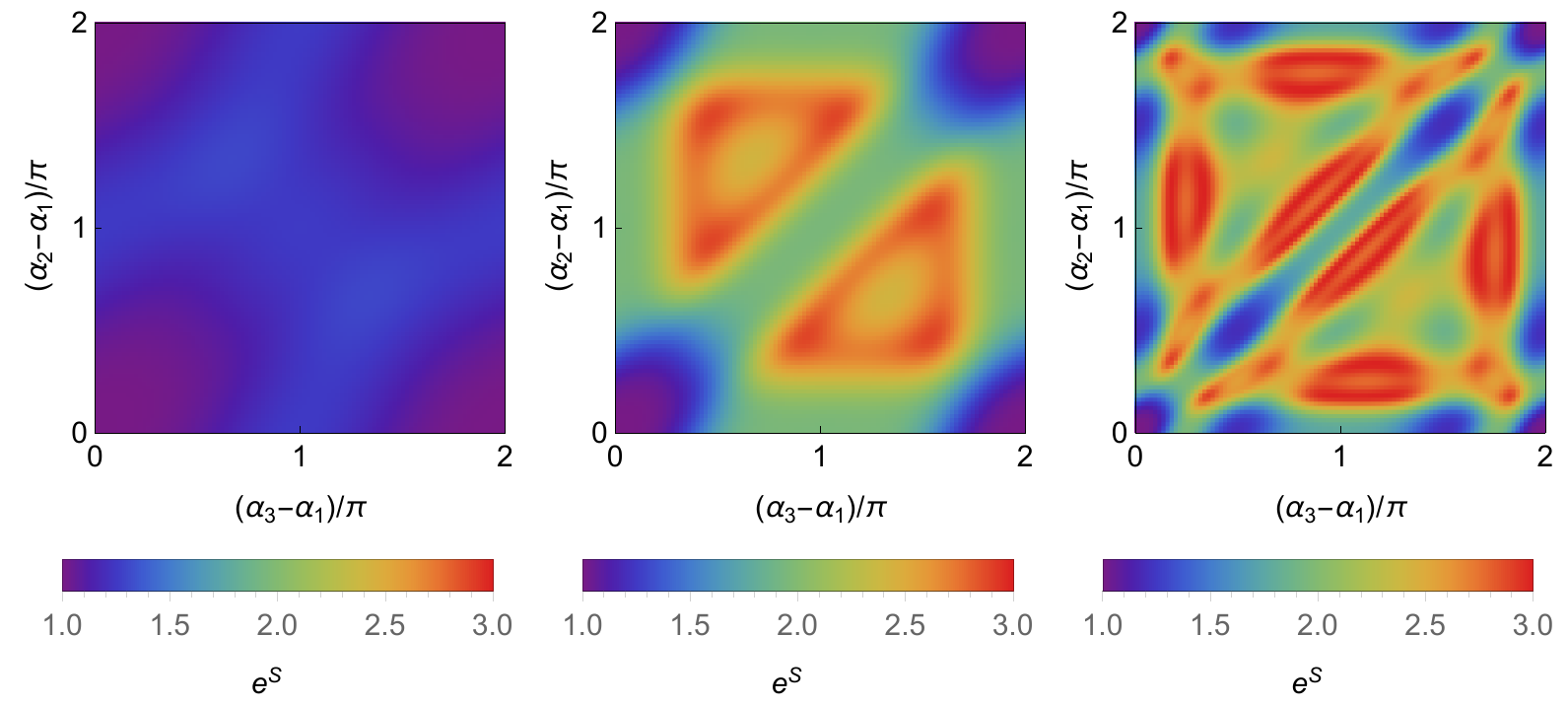}
\caption{Color plots of $\e^S$, where $S$ is the entanglement entropy, for the pair of photons reflected from three modulated qubits as a function of the  relative phases of the modulation $\alpha$ calculated for different modulation amplitudes (from left to right): $A/\Omega = 1$, $2$, and $5$.
}\label{fig:entr3}
\end{figure}


The approach can be generalized to the larger number of qubits $N$ modulated with phases $\alpha_a$, $a=1,2,...,N$.  The wave function of the emitted photon pair calculated using the same approach reads
\begin{align}
    \psi_{n_1,n_2} \propto  \sum_{a \neq b} a_{n_1}^{(a)}a_{n_2}^{(b)} \,.
\end{align}
It is instructive to rewrite it as
\begin{align}
    \psi_{n_1,n_2} \propto   
     \left(\sum_a a_{n_1}^{(a)} \right) \left( \sum_a a_{n_2}^{(a)} \right)- \sum_a  a_{n_1}^{(a)}a_{n_2}^{(a)} \,.
\end{align}
This expression features $N+1$ terms that are all  factorized with respect to $n_1$ and $n_2$. However, only $N$ of the corresponding eigenvectors are linearly independent. Therefore, the entanglement entropy cannot exceed $\ln N$. 

Figure~\ref{fig:entr2} shows the calculated entanglement entropy as a function of the relative modulation phases $\alpha_2-\alpha_1$ and $\alpha_3-\alpha_1$ for different modulation amplitudes $A$. As predicted, the entropy vanishes at zero relative phases. To achieve the maximally possible value $S=\ln 3$, the easiest way is to choose $\alpha_1-\alpha_2 = \alpha_2-\alpha_3 = 2\pi/3$, so that the three scalar products $(a^{(1)},a^{(2)})$, $(a^{(2)},a^{(3)})$, $(a^{(3)},a^{(1)})$ are equal. Then, it easy to check that the state 
\begin{align}
    \psi_{n_1,n_2} = a^{(1)}_{n_1} a^{(2)}_{n_2} +a^{(2)}_{n_1} a^{(1)}_{n_2} +
    a^{(2)}_{n_1} a^{(3)}_{n_2} +a^{(3)}_{n_1} a^{(2)}_{n_2} +
    a^{(3)}_{n_1} a^{(1)}_{n_2} +a^{(1)}_{n_1} a^{(3)}_{n_2}
\end{align}
will have three equal singular values if $(a^{(1)},a^{(2)})=(a^{(2)},a^{(3)})=(a^{(3)},a^{(1)})=-1/5$, which leads to the condition
\begin{align}
    J_0\left(\frac{\sqrt{3}A}{\Omega} \right) = -\frac15 \,.
\end{align}
The smallest amplitude which fulfills the condition is $A/\Omega \approx 1.64$. The wave function of the photon pair in that case has the form
\begin{align}
    \psi_{n_1,n_2} = \frac1{\sqrt{3}}(u_{n_1}u_{n_2} + v_{n_1}v_{n_2} + w_{n_1}w_{n_2})
\end{align}
which is a Bell state of a pair of qutrits~\cite{Migdal2013} (three-level qudits) with the basis states $(u, v, w)$. As mentioned previously, this basis can be easily changed using pulse shapers and linear modulators. We note also, that in the case of high cooperativities, when the qubit predominantly relaxes to the waveguide mode, the discussed protocol allows for the deterministic generation of the Bell state of the qudit pairs, which is of paramount importance for the development of the quantum information processing. Importantly, the frequency bins for the qudit states can be dynamically tuned by the driving frequency unlike the set-ups involving strong coupling of qubits to the cavity modes where the frequency bins are determined by the Rabi frequency and therefore by the set up geometry. Moreover, the number of frequency bins (sidebands) in our system is not limited by 2 (as it is in case of Rabi doublet), so the much more complicated many-qudit states, as described in Sec. S7, can be generated.

\section{Designing correlations}

Here we show that non-Harmonic modulation of the qubits can be used to realize a desired two-photon state.  
For two modulated qibits, the wave function of the scattered photon pair  can be calculated from Eq.~\eqref{eq:psinn}. 
Suppose we have  a target two-photon wave function $\Psi_{n_1,n_2}$ that we want to obtain (up to a constant factor) by inducing modulation of a specific shape. To this end, we follow the algorithm:

\begin{enumerate}
    \item Do an SVD factorization of the target wave function:
    \begin{align}
        \psi_{n_1,n_2} \approx  \lambda^{(1)} \alpha_{n_1}^{(1)} \alpha_{n_2}^{(1)} +
         \lambda^{(2)} \alpha_{n_1}^{(2)} \alpha_{n_2}^{(2)}
    \end{align}
    If the SVD decomposition  has more than 2 terms, more than 2 qubits are required to realize it. 
    
    \item Define 
    \begin{align}
        a_n^{(1,2)} = C^{(1,2)} \left[\sqrt{\lambda^{(1)}}\alpha_{n}^{(1)} \pm i \sqrt{\lambda^{(2)}}\alpha_{n}^{(2)} \right] \,,
    \end{align}
    where the normalization constants $C^{(1,2)}$ are chosen to ensure $\sum_n |a_n^{(1,2)}|^2 =1$. Then, 
     $
        \psi_{n_1,n_2} \approx 2 C^{(1)}C^{(2)}\left(a_{n_1}^{(1)} a_{n_2}^{(2)} + a_{n_1}^{(2)} a_{n_2}^{(1)} \right)
    $.
    
    \item Calculate required modulation of the qubit resonance frequencies
    \begin{align}
    \omega^{(1,2)}(t) = \rmi \frac{d}{dt} \ln \left[\sum_n a_n^{(1,2)} e^{-i n \Omega t} \right] \,.
    \end{align}
    In general case, this expression has both imaginary and real parts, i.e., both the resonance frequency and the decay rate should be modulated. 
\end{enumerate}

As an example, suppose we want to realize a Bell state
\begin{align}
    \Psi_{n_1,n_2} =\frac1{\sqrt2} (f_{n_1} f_{n_2} + g_{n_1} g_{n_2}) \,,
\end{align}
where
$f_n = \delta_{n,0}$ and $g_n = (\delta_{n,1} + \delta_{n,-1})/\sqrt{2}$. Following the above procedure, we arrive to the required modulation,
\begin{align}
    \omega^{(1,2)}(t) = {\rm Im\,}\frac{\rmd}{\rmd t} \ln (1 \pm  \rmi\, \sqrt2 \cos \Omega t) \,,
\end{align}
where we disregarded imaginary part of the modulation. Despite that, the target wave function is reproduced with the fidelity $F = 0.9$,
\begin{align}
    F = \left|\sum_{n_1,n_2} \Psi_{n_1,n_2}^{*} \psi_{n_1,n_2} \right|^2 \,,
\end{align}
where both $\Psi_{n_1,n_2}$ and $\psi_{n_1,n_2}$ are supposed to be normalized. It should be noted, that in order to further increase the fidelity, one would require modulate both real and imaginary component of the qubit frequency: ultimately to achieve the fidelity equal to unity, would require modulation which would change the sign of the decay rate and thus would require amplification. At the same as can be seen the purely real frequency modulation allows to have fidelity 0.9. Further improvements may be achieved by taking larger number of qubits and using the numerical optimization with additional constraint of purely real modulation to maximize fidelity. 

\blue{
\section{Correlations of $M \geq 3$ photons}

\begin{figure}[t]
\label{fig:entr33}
\centering
\includegraphics[width=0.99\textwidth]{ 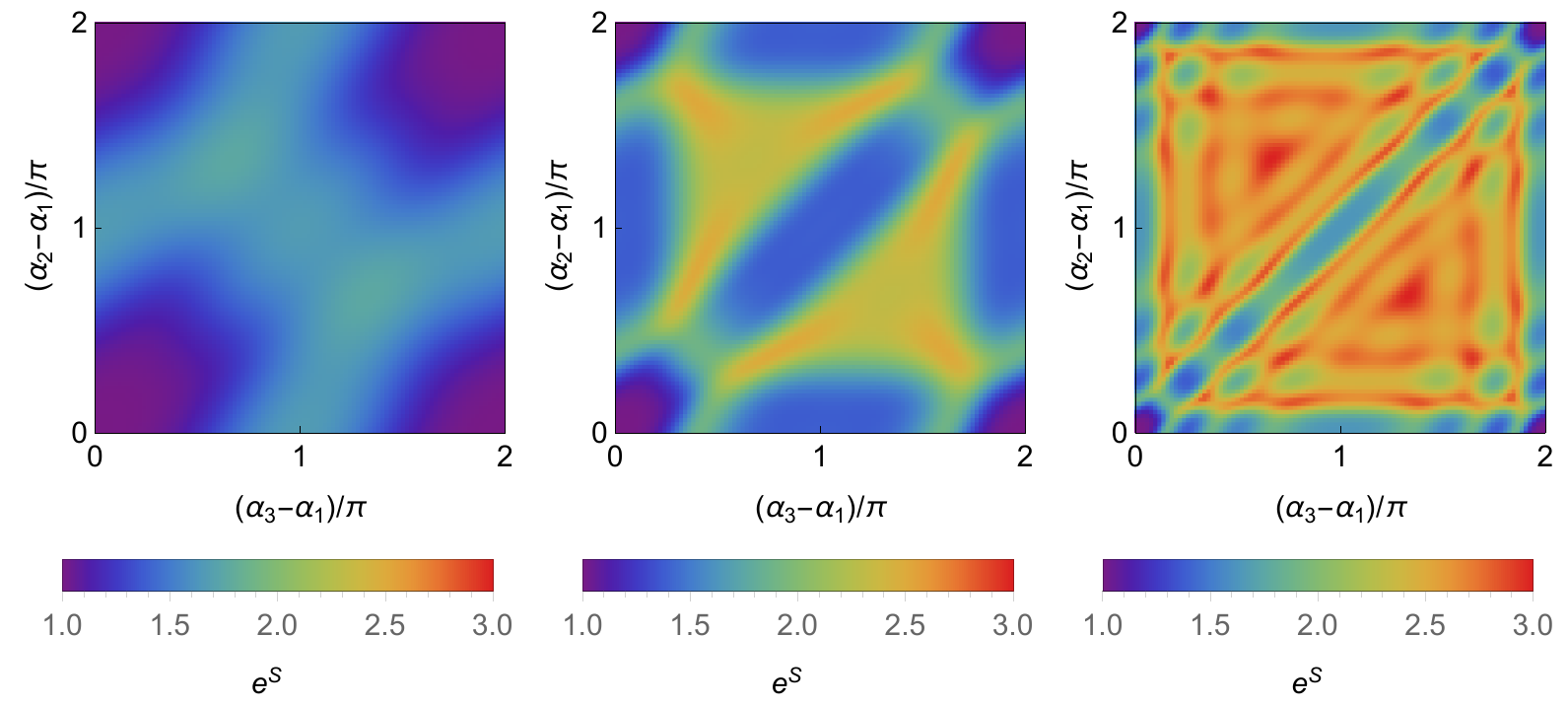}
\caption{Color plots of $\e^S$, where $S$ is the entanglement entropy, for the three photons reflected from three modulated qubits as a function of the  relative phases of the modulation $\alpha$ calculated for different modulation amplitudes (from left to right): $A/\Omega = 1$, $2$, and $5$.
}\label{fig:entr33}
\end{figure}


Here we demonstrate that the system with $N \geq M$ 
modulated qubits can generate entangled $M$-photons states. There exists no conventional universal measure of entanglement for $M > 2$ particles, and here  we chose  as such the entanglement entropy based on higher-order singular value decomposition, defined according to Ref.~\cite{Poshakinskiy2021}. 
}

We start with the three-photon states.
Generalizing the  approach of Sec.~\ref{sec:simple}, we calculate the three-photon wave function as
\begin{align}
   \psi_{n_1,n_2,n_3} =  \sum_{a \neq b, b \neq c, c \neq a }a_{n_1}^{(a)} a_{n_2}^{(b)} a_{n_3}^{(c)} 
\end{align}
where $a_n^{(i)}$ is defined by Eq.~\eqref{eq:an}. 
The entanglement entropy is evaluated  as follows~\cite{Poshakinskiy2021}: We calculate the  higher-order singular value decomposition of the wave function
\begin{align}\label{eq:hosvd}
    \psi_{n_1,n_2,n_3} = \sum \Lambda_{\nu_1 \nu_2 \nu_3} U^{\nu_1}_{n_1} U^{\nu_2}_{n_2} U^{\nu_3}_{n_3} \,,
\end{align}
define the singular values as
\begin{align}\label{eq:holambda}
    |\lambda_\nu|^2 = \sum_{\nu_1 \nu_2} |\Lambda_{\nu_1 \nu_2 \nu}|^2 \,,
\end{align}
normalize them and use the definition of the entropy Eq.~\eqref{eq:defentr}.

The result for $N=3$ qubits  is shown in Fig.~\ref{fig:entr33}. Similarly to the two-photon entropy, the three-photon entropy is limited by $\ln 3$. This maximal value is reached when the three
frequency combs generated by the three qubits are orthogonal, i.e,
Eq.~\eqref{eq:entr2} must be fulfilled simultaneously for $\alpha=\alpha_1-\alpha_2$, $\alpha_2-\alpha_3$, and $\alpha_1-\alpha_3$. That is realized at smallest possible amplitude if we choose $\alpha_1-\alpha_2=  \alpha_2-\alpha_3 = 2\pi/3$ and $A/\Omega= j_{0,1}/\sqrt{3} \approx 1.39$, where $j_{0,1}$ is the first zero of the Bessel function $J_0$. The resulting state reads
\begin{align}\label{eq:cluster}
   \psi_{n_1,n_2,n_3}^{\rm (cl)} =  a_{n_1}^{(1)} a_{n_2}^{(2)} a_{n_3}^{(3)} +
   a_{n_1}^{(1)} a_{n_2}^{(3)} a_{n_3}^{(2)} +
   a_{n_1}^{(2)} a_{n_2}^{(1)} a_{n_3}^{(3)} +
   a_{n_1}^{(2)} a_{n_2}^{(3)} a_{n_3}^{(1)} +
   a_{n_1}^{(3)} a_{n_2}^{(1)} a_{n_3}^{(2)} +
   a_{n_1}^{(3)} a_{n_2}^{(2)} a_{n_3}^{(1)}\:. 
\end{align}

The state  $\psi^{\rm (cl)}$ can be regarded as a generalization of the 3-qubit cluster state for the case of many-level qudits.
\blue{ Indeed, the main property of the usual cluster states of qubits is that if one of the qubits is measured, the other remain in the entangled cluster state. Based on this, the cluster quantum computing is realized.  Similarly, for the qudit state $\psi^{\rm (cl)}$, 
 if one of the photons is measured in the basis $(a^{(1)},a^{(2)},a^{(3)})$, the other two remain to be entangled and form the Bell state, e.g., if the first qubit was measured in the $a^{(1)}$ state, the remaining two are in the state $\psi_{n_2,n_3} \propto a_{n_2}^{(3)} a_{n_3}^{(2)} + a_{n_2}^{(2)} a_{n_3}^{(3)}$. 
 }
However, note that the  state Eq.~\eqref{eq:cluster} is different from  the three-qutrit analogue GHZ state  $\psi^{\rm GHZ}_{n_1,n_2,n_3} = s_{n_1}s_{n_2}s_{n_3}+t_{n_1}t_{n_2}t_{n_3}+u_{n_1}u_{n_2}u_{n_3}$~\cite{Migdal2013}, i.e., cannot be reduced to it by a local  basis transformation.
However, both states have equally high entanglement entropy $\ln 3$.

\blue{

\begin{figure}[t]
\label{fig:entr3N}
\centering
\includegraphics[width=0.6\textwidth]{ 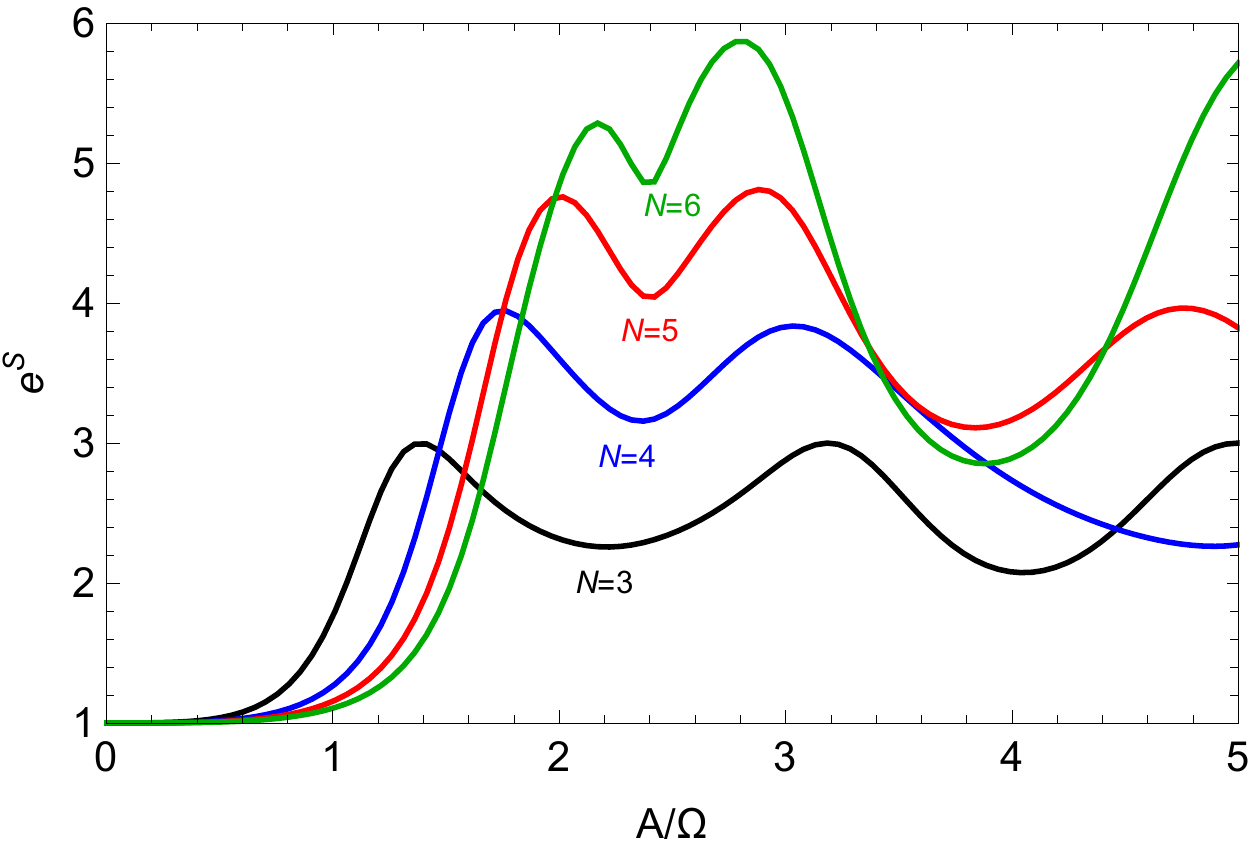}
\caption{
Dependence of $\e^S$, where $S$ is 3-photon the entanglement entropy, for the three photons reflected from $N$ modulated qubits as a function of the  modulation amplitude $A$ for different $N$. The qubits are modulated with the phases $\alpha_i = 2\pi i/N$ and equal amplitudes.
}\label{fig:entr3pN}
\end{figure}


\begin{figure}[t]
\label{fig:entr4N}
\centering
\includegraphics[width=0.6\textwidth]{ 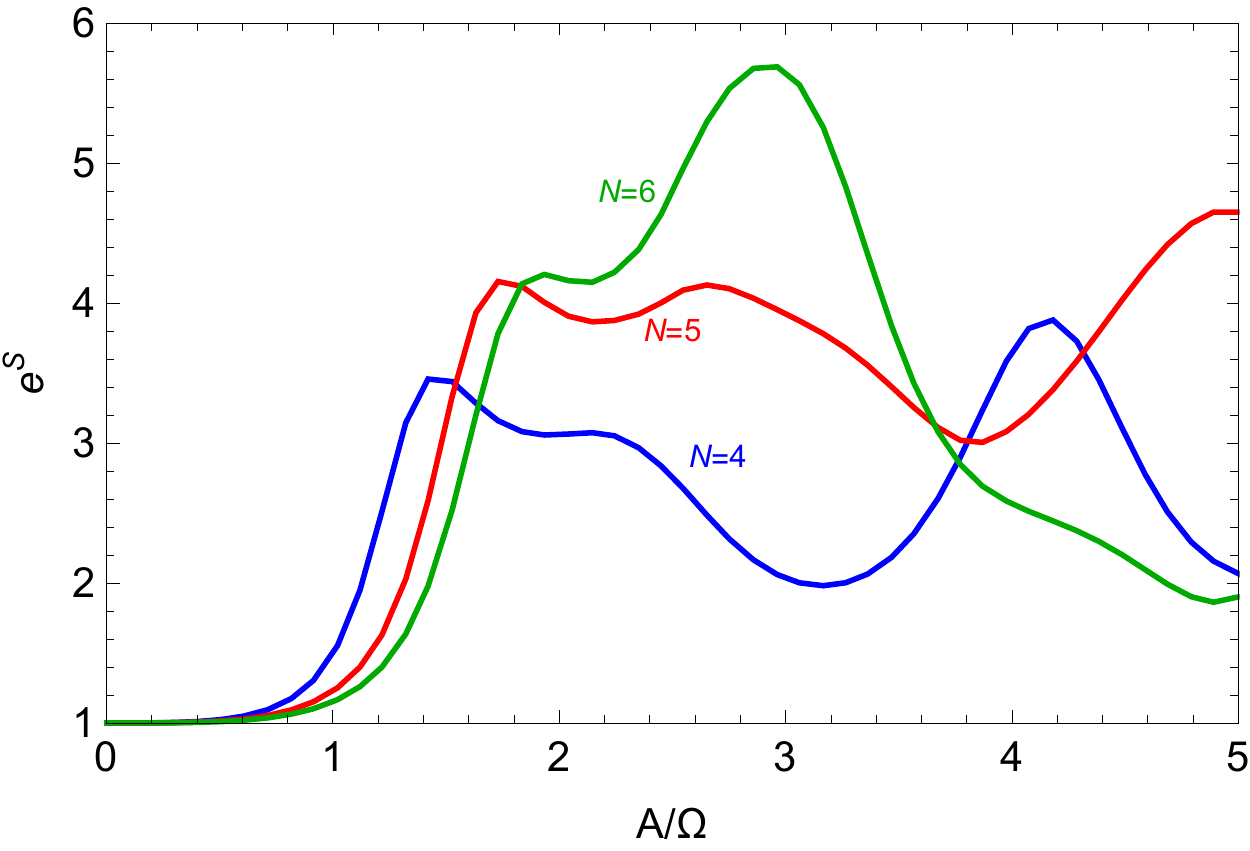}
\caption{
Dependence of $\e^S$, where $S$ is 4-photon the entanglement entropy, for the three photons reflected from $N$ modulated qubits as a function of the  modulation amplitude $A$ for different $N$. The qubits are modulated with the phases $\alpha_i = 2\pi i/N$ and equal amplitudes.
}\label{fig:entr4pN}
\end{figure}


Three-photon states with higher entanglement entropy can be realized in the system with larger number of modulated qubits. Indeed, the three-photon state generated by $N>3$ qubits lays in the Hilbert space spanned by the $N$ basis vectors $a_n^{(i)}$, defined by Eq.~\eqref{eq:an}, so the entanglement entropy cannot exceed $\ln N$. 
Fig.~\ref{fig:entr3pN}
the entanglement entropy as a function of modulation amplitude $A$ for $N=3,4,5,6$ qubits modulated with equal amplitudes $A$ and different phases $\alpha_i = 2\pi i/N$ ($i=1, 2, ..., N$). 
One can see, that at certain values of amplitude $A$, the entropy approaches its upper bound $\ln N$.

All the above results remain valid for 4-photon states emitted by $N \geq 4$ modulated qubits. Fig.~\ref{fig:entr4pN} shows the calculated
the entanglement entropy as a function of modulation amplitude $A$ for $N=4,5,6$. The entropy is bound by the same value $\ln N$. 

Finally, we consider $M$-photon states generated by $N=M$ modulated qubits and show how the highest possible entanglement entropy $S= \ln N$ can be achieved. For $N=M$, the $M$-photon wave function has the form 
\begin{align}\label{eq:psiMM}
   \psi_{n_1,n_2,...,n_N} = \frac1{\sqrt{N!}}\sum_{(k_1,k_2,...,k_N)} a_{k_1}^{(1)} a_{k_2}^{(2)} ... a_{k_N}^{(N)} \,,
\end{align}
where $a_n^{(i)}$ is defined by Eq.~\eqref{eq:an} and the sum is taken over all  permutations $(k_1,k_2,...,k_N)$ of the indices $(1,2,...,N)$. 
Suppose now that all $a_n^{(i)}$ are mutually orthogonal. Then, the state Eq.~\eqref{eq:psiMM} is the $M$-photon generalization of the 3-qutrit  state Eq.~\eqref{eq:cluster}. 
Its higher-order singular value decomposition is easily performed. Indeed, if we choose in the decomposition definition Eq.~\eqref{eq:hosvd} $U_n^\nu = a_n^{(\nu)}$, we arrive to the core matrix $\Lambda_{\nu_1,\nu_2,...,\nu_N}$ with the elements equal to $1/\sqrt{N}$ when the indices $\nu_1\nu_2...\nu_N$ are all different. Then, $|\lambda_\nu|^2$, calculated according to Eq.~\eqref{eq:holambda}, are all equal $(1/N!)\cdot (N-1)\cdot(N-2)\cdot...\cdot 1 = 1/N$. Thus, $S = \ln N$ which is the highest possible value of the entanglement entropy. 

\begin{figure}[t]
\label{fig:entrX}
\centering
\includegraphics[width=0.6\textwidth]{ 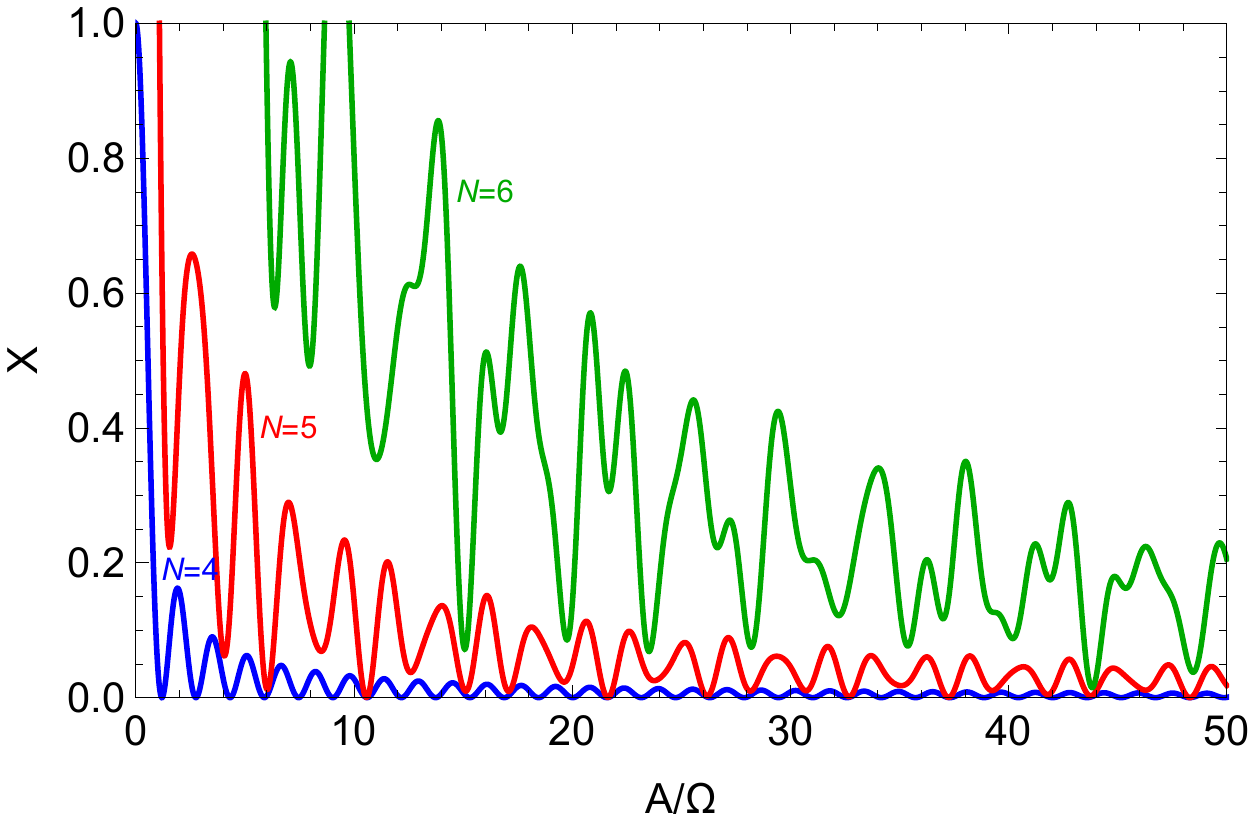}
\caption{
Dependence of $X$, that is calculated according to Eq.~\eqref{eq:X} and quantifies the deviation of the entanglement entropy from its maximal value, of the modulation modulation amplitude $A$ for the cases of $N=2,4,8$ qubits. The qubits are modulated with the phases  $\alpha_i = 2\pi i/N$ and equal amplitudes $A$.
}\label{fig:entrX}
\end{figure}


Therefore, to generate the maximally entangled state, the modulations of all the $N$ qubits must be mutually orthogonal in terms of the scalar product~Eq.\eqref{eq:scal}. These $N(N-1)/2$ conditions, for large $N$, cannot be fulfilled by tuning $2N$ parameters -- the amplitudes and phases 
of the harmonic modulation of $N$ qubits.  Therefore, non-harmonic modulations are required in general case. However, the orthogonality can be achieved approximately for harmonic modulation with arbitrarily high precision provided the modulation amplitude $A$ is high enough. Figure~\ref{fig:entrX} shows the sum of all scalar products
\begin{align}\label{eq:X}
   X= \sum_{i<j}  |(a^{(i)},a^{(j)})|^2 = \sum_{i<j} J_0^2\left(\frac{2A}{\Omega} \, \sin \frac{\alpha_i-\alpha_j}{2} \right)\,,
\end{align}
that quantifies the deviation of the entanglement entropy from its maximal value, as the function of harmonic modulation amplitude for the cases of $N=2,4,8$ qubits. The Bessel functions in the sum Eq.~\eqref{eq:X} oscillate with incommensurate periods. So, for high enough $A$, their zeroes can occur at arbitrarily close points. There,  $X$ almost vanishes, meaning that the modulations  $a^{(i)}$ are almost orthogonal. The larger is $N$, the higher value of $A$ is required to achieve the orthogonality of a certain precision. }

\bibliography{trembling}

\end{document}